\documentclass[graybox]{svmult}

\usepackage{type1cm}        % activate if the above 3 fonts are
\usepackage{makeidx}         % allows index generation
\usepackage{graphicx}        % standard LaTeX graphics tool
\usepackage{multicol}        % used for the two-column index
\usepackage[bottom]{footmisc}% places footnotes at page bottom
\usepackage{url}

\usepackage{newtxtext}       % 
\usepackage[varvw]{newtxmath}       % selects Times Roman as basic font
\usepackage[colorlinks=true,allcolors=blue]{hyperref}

\makeindex             % used for the subject index
\begin{document}

\title*{Microscopy of Ultracold Fermions in Optical Lattices}

\author{Waseem Bakr, Zengli Ba and Max Prichard}
\institute{Waseem Bakr \at Princeton University, Dept. of Physics, Princeton University, Princeton, New Jersey 08544, USA, \email{wbakr@princeton.edu}
\and Zengli Ba \at Princeton University, Dept. of Physics, Princeton University, Princeton, New Jersey 08544, USA, \email{zlba@princeton.edu}
\and Max Prichard \at Princeton University, Dept. of Physics, Princeton University, Princeton, New Jersey 08544, USA, \email{mprichard@princeton.edu}}

%\and Name of Second Author \at Name, Address of Institute \email{name@email.address}}
\maketitle

\abstract{These lecture notes review recent progress in studying the Fermi-Hubbard model using ultracold gases in optical lattices. We focus on results from quantum gas microscope experiments that have allowed site-resolved measurements of charge and spin correlations in half-filled and doped Hubbard systems, as well as direct imaging of various types of polaronic quasiparticles. We also review experiments exploring dynamical properties of the Hubbard model through transport and spectroscopy. Moving beyond the plain-vanilla square-lattice Hubbard model, we present more recent work exploring Hubbard systems with novel lattice geometries and long-range interactions that stabilize new phases. Finally, we discuss the realization of entropy distribution protocols to cool these systems to ultralow temperatures where comparison to unbiased numerics is no longer possible.}

\section{Introduction}
\label{sec:1}

\subsection{Studying many-body physics with ultracold gases}

Strongly correlated quantum materials are at the forefront of current condensed matter research~\cite{coleman2015introduction, fulde2012correlated}. They exhibit low-temperature phases with fascinating properties, as well as unusual quasiparticle excitations and dynamics. Examples include high-temperature superconductors, quantum magnets, fractional quantum Hall systems, and quantum spin liquids. Understanding the microscopic physics of these quantum systems is useful for engineering their properties for technological applications. However, simulations on classical computers quickly run into the many-body problem, limiting exact techniques to studying a few dozen particles. Quantum simulators present a way to circumvent this problem, and are being pursued using a variety of platforms, including cold atoms, ions, photons and superconducting qubits~\cite{feynman1982simulating,georgescu2014quantum}. 

Degenerate quantum gases are a particularly fruitful platform for simulating many-body Hamiltonians that describe interacting electrons or spins~\cite{bloch2008many,bloch2012quantum,gross2017quantum}. They offer the experimentalist a high degree of control in tailoring these Hamiltonians and controlling their parameters, often in real time. Compared to condensed matter systems, the microscopic Hamiltonians that describe cold atom systems are much better understood from first principles, allowing benchmarking against state-of-the-art numerics in regimes where it is feasible. The strongly interacting regime can be achieved using optical lattices, which quench the kinetic energy, or by tuning close to Feshbach resonances which directly control the interactions between the atoms. 

Being composite particles, atoms can be used to study bosonic or fermionic systems. In these lectures, we will focus on Fermi gases in optical lattices, which can be thought of as enlarged models for itinerant electronic materials, with the atoms playing the role of the electrons and the optical lattice playing the role of the ionic lattice in a solid. With this picture in mind, two additional advantages of cold atoms become apparent. The micron-scale spacing and heavy mass of the atoms compared to electrons lead to dynamics on the millisecond scale (as opposed to femtosecond dynamics in solids), which enables the cold atom experimentalist to easily study non-equilibrium dynamics. Furthermore, the large spacing opens up the possibility of optically probing and manipulating cold atom systems at the smallest relevant length-scale, corresponding to the lattice spacing. This capability is realized using the technique of quantum gas microscopy~\cite{gross2021quantum}, first introduced in 2009 for bosonic gases~\cite{bakr2009quantum,bakr2010probing,sherson2010single}, and later extended to Fermi gases in 2015~\cite{parsons2015site, cheuk2015quantum,omran2015microscopic,haller2015single,edge2015imaging} .

\subsection{The Fermi-Hubbard model}
Ultracold fermions in an optical lattice provide a clean realization of the Fermi-Hubbard model. This model Hamiltonian was introduced in 1963 in independent papers by Gutzwiller, Kanamori and Hubbard~\cite{gutzwiller1963effect,kanamori1963electron,hubbard1963electron}. The original motivation for introducing it was to explain itinerant ferromagnetism in transition metals. Following the discovery of the high-temperature superconducting cuprates by Bednorz and M\"{u}ller in 1986 \cite{bednorz1986possible}, Anderson proposed the Hubbard model as a minimal model to capture the key physics of these materials~\cite{anderson1987resonating}. This motivated much theoretical work on the Hubbard model, whose scope of application has expanded greatly since, and is now regarded as a paradigmatic model for studying strong correlation physics of itinerant lattice fermions. 

Despite decades of theoretical work, the Hubbard model has been solved exactly only in the 1D case using the Bethe ansatz~\cite{lieb1968absence}. In higher dimensions, particularly the 2D case relevant to the cuprates, much effort has been devoted to developing numerical techniques to understand its low-temperature phase diagram~\cite{leblanc2015solutions,schafer2021tracking}. Methods based on exact diagonalization are limited to lattices with about 20 sites. Quantum Monte Carlo (QMC) methods, such as determinantal QMC, provide a powerful and unbiased approach for calculating correlation functions on larger lattices, but the infamous fermion sign problem leads to an exponential computational barrier at low temperatures~\cite{loh1990sign}. Other techniques are biased towards particular wavefunctions or are approximate. For example, a popular technique is Dynamical Mean Field Theory (DMFT), which maps the lattice problem to a single or multi-site impurity problem that is more computationally tractable. However, DMFT becomes exact only in the infinite coordination limit~\cite{georges1996dynamical}. These limitations in accessing the low-temperature equilibrium and dynamical properties of the Fermi-Hubbard model using numerics motivate quantum simulations with cold atoms. Besides the potential to reveal novel emergent phenomena in regimes outside the reach of unbiased numerics, these experiments also provide data to benchmark theoretical techniques that are widely used in other contexts in the field of condensed matter physics.

The ``plain-vanilla" Fermi-Hubbard Hamiltonian is given by
\begin{equation}
\label{eq:hubbard_model}
\hat{H}=-t\sum_{\langle i,j\rangle,\sigma\in\{\uparrow,\downarrow\}}(\hat{c}_{i,\sigma}^{\dagger}\hat{c}_{j,\sigma} + \mathrm{h.c.})+U\sum_i\hat{n}_{i\uparrow}\hat{n}_{i\downarrow},
\end{equation}
where $\hat{c}^{\dagger}_{i\sigma} $ ($\hat{c}_{i\sigma}$) creates (destroys) a fermion of spin $\sigma$ at lattice site $i$, $\hat{n}_{i\sigma} = \hat{c}^{\dagger}_{i\sigma}\hat{c}_{i\sigma}$ is the number operator measuring the number of fermions of spin $\sigma$ on site $i$, $\langle i,j\rangle$ denotes nearest-neighbor sites, $t$ is the nearest-neighbor tunneling rate and $U$ in the on-site interaction. The dimensionless parameters of the model are the coupling strength $U/t$, the temperature $T/t$ (setting the Boltzmann constant $k_B = 1$) and the doping $\delta = n - 1$, where $n$ is the average number of fermions on a lattice site. When the number of fermions and sites is equal, the system is said to be at ``half-filling" or undoped ($\delta=0$). Hole (particle) doping corresponds to $\delta < 0$ ($\delta > 0$). 

The single-band model described by Eq.~\ref{eq:hubbard_model} is commonly used on the square lattice as a very simplified model of the cuprate superconductors~\cite{anderson1997theory}. The cuprates are layered materials, where the superconductivity is thought to arise from electrons moving in the copper oxide planes. Intervening layers of other metal oxides act as charge reservoirs that dope the copper oxide planes. The low-energy electronic physics is determined by the Cu 3$d_{x^2-y^2}$ and O 2$p_x$, 2$p_y$ orbitals. This suggests writing down a 2D three-band model focusing on the copper oxide planes, which is indeed used sometimes, but it can be further simplified to the effective single-band model of Eq.~\ref{eq:hubbard_model}  as shown by Zhang and Rice~\cite{zhang1988effective}.  

\begin{figure}[t]

%\sidecaption
% For example, with the graphicx style use
\centering
\includegraphics[width=10cm]{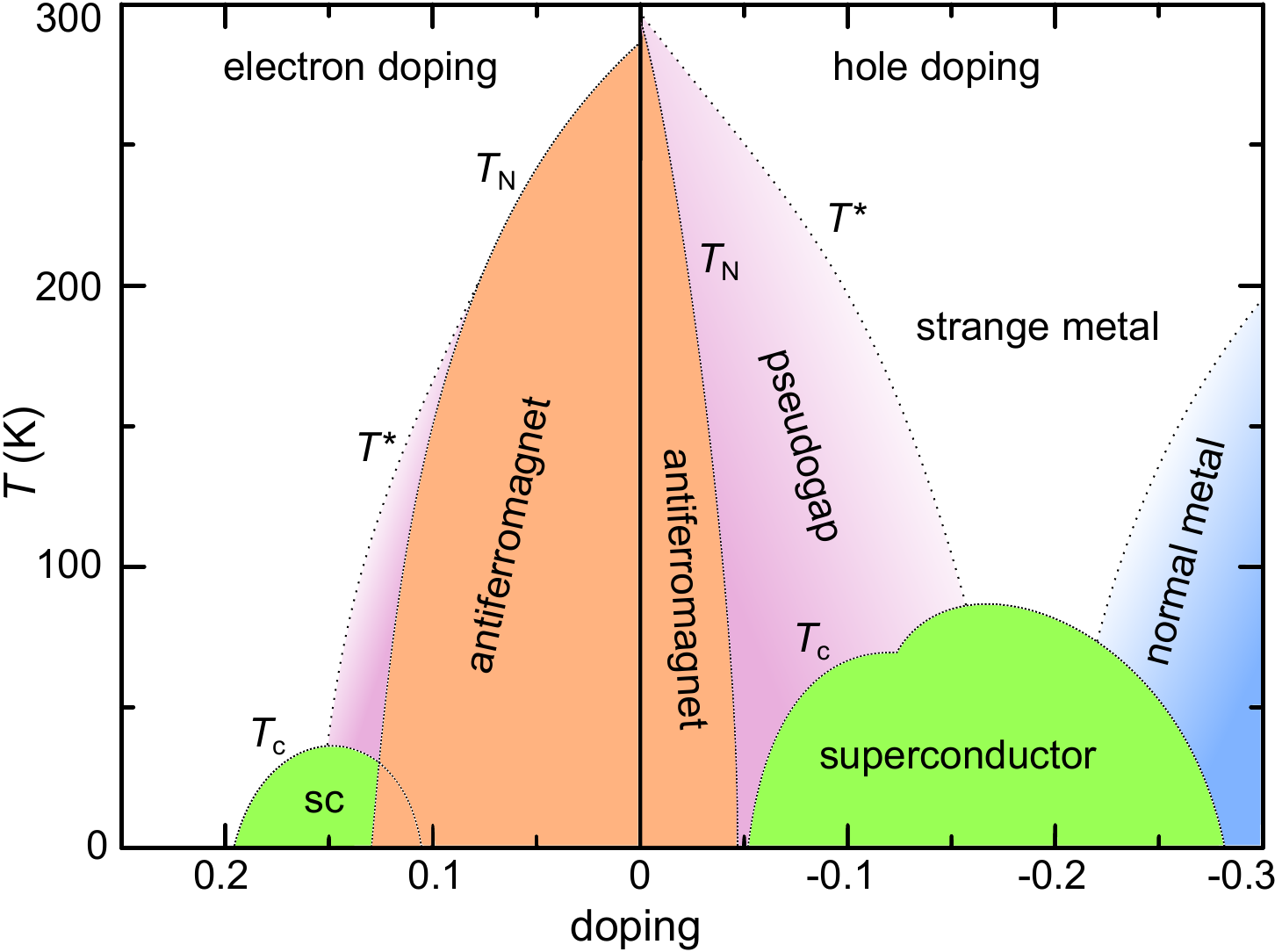}
%
% If no graphics program available, insert a blank space i.e. use
%\picplace{5cm}{2cm} % Give the correct figure height and width in cm
%
\caption{Cartoon phase diagram of the high-temperature superconducting cuprates. Adapted from~\cite{motzkau2013}.}
\label{fig:cuprates}       % Give a unique label
\end{figure}

A cartoon phase diagram of the cuprates is shown in Fig.~\ref{fig:cuprates}, highlighting some of the key experimentally observed phases. At half-filling, where each copper atom has a single electron in its $d_{x^2-y^2}$ orbital, the cuprates are antiferromagnetic Mott insulators. In the Mott insulator, the on-site Coulomb repulsion is strong compared to the electron tunneling. This suppresses double occupancies on a site and leads to an energy gap that inhibits electron transport. The antiferromagnetic order is suppressed upon doping with electrons or holes. At a certain value of doping, the insulator transitions into a superconductor whose pairing gap exhibits a $d$-wave symmetry. The superconductor is characterized by a high critical temperature and a short coherence length. It is called an unconventional superconductor as it cannot be described using the well-developed Bardeen-Cooper-Schrieffer (BCS) theory that describes superconductivity in materials like aluminum or niobium. The normal phase of the cuprates also exhibits unusual properties. Interesting regimes include the pseudogap regime, where a spectral gap opens up near the Fermi surface, and the strange metal, which exhibits anomalous transport properties. A key theoretical question in the cuprates is understanding the ``pairing glue" that leads to Cooper pairing despite the repulsive interactions between the electrons. 

The Fermi-Hubbard model is known to reproduce some of the phenomenology of the cuprates. At half-filling, QMC does not have a fermion sign problem and calculations can be done down to very low temperatures. These numerics confirm the antiferromagnetic Mott insulator as the ground state. However, the sign problem makes numerical studies of the doped part of the diagram very challenging, and the nature of the ground state has been the subject of much debate. State of the art calculations favor stripe order as the ground state rather than superconductivity, but superconductivity may be stabilized by extending the model with a small diagonal tunneling (usually denoted as $t'$)~\cite{zheng2017stripe,qin2020absence,xu2024coexistence}. This diagonal tunneling also breaks the particle-hole symmetry of the Hamiltonian in Eq.~\ref{eq:hubbard_model}, which is clearly absent in the phase diagram in Fig.~\ref{fig:cuprates}.

\section{Experimental techniques}
The starting point for studying the Hubbard model with cold atoms is the production of a degenerate Fermi gas by laser cooling followed by evaporative cooling. The atomic species typically used in these experiments are fermionic alkalis, namely $^6$Li or $^{40}$K, because of their simple level structure, which makes them particularly amenable to laser cooling. For evaporative cooling, the atoms are prepared in an optical trap in an equal mixture of two hyperfine states. The requirement of wavefunction anti-symmetrization for two identical colliding fermions only allows collisions in odd partial waves, but those are frozen out at ultracold temperatures, rendering a spin-polarized Fermi gas non-interacting. On the other hand, fermions in different internal states collide in the $s$-wave channel, enabling evaporative cooling to quantum degeneracy. The two hyperfine states also serve to encode the pseudospin-1/2 in the realization of the Hubbard model.

\subsection{Quantum gas microscopy}
\label{sec:quantumg_gas_microscopy}

Quantum gas microscopy is a technique for probing and manipulating 2D Hubbard systems in optical lattices with single-atom and single-site resolution~\cite{gross2021quantum}. Starting from a weakly correlated state, namely a continuum Fermi gas, an optical lattice is adiabatically introduced to quench the kinetic energy and enhance the effect of interactions. The resulting strongly correlated state is then frozen by rapidly ramping up the lattice depth, fast compared to the microscopic Hubbard timescales (tunneling and interaction times), but slowly compared to the lattice band gap to avoid excitation to higher bands. The system is then probed with fluorescence imaging, which projects the complicated many-body state onto a particular realization in the measurement basis. By repeatedly preparing and measuring the system under identical conditions, various $n$-point correlations of spin and density can be extracted, providing a microscopic characterization of the many-body state. Another class of experiments involves locally manipulating the system by flipping spins or introducing potential perturbations that modify the density on a small length scale (e.g., inserting a single hole)~\cite{weitenberg2011single,ji2021coupling}. The dynamics of these excitations can then be probed by repeatedly preparing the system and evolving for different times before taking the fluorescence snapshots.

The first quantum gas microscopes were introduced for $^{87}$Rb, a bosonic species, by the Harvard and MPQ groups in 2009~\cite{bakr2009quantum,bakr2010probing,sherson2010single}. They were used to study many phenomena in the Bose-Hubbard model, including number squeezing across the superfluid to Mott insulator transition, high-resolution imaging of shell structure in the bosonic Mott insulator and quantum magnetism in tilted lattices~\cite{bakr2010probing,sherson2010single,simon2011quantum}. Recognizing the utility of the technique, several groups started construction of microscopes for fermionic species, with the first papers reporting images taken with these microscopes appearing in 2015. The Harvard \cite{parsons2015site}, MPQ \cite{omran2015microscopic} and Princeton \cite{brown2017spin} groups used $^6$Li, while the MIT \cite{cheuk2015quantum}, Toronto \cite{edge2015imaging} and Strathclyde \cite{haller2015single} groups used $^{40}$K. 

All quantum gas microscope experiments share three key technical elements. First, a high-resolution objective is needed, typically with a numerical aperture (NA) above 0.5. This is required to obtain an almost diffraction-limited point-spread-function size, comparable to or below the lattice spacing. To maintain fast tunneling for adiabatic state preparation, the lattice spacing is typically $\lesssim0.75~\mu$m. The large numerical aperture also enables large photon collection efficiencies for improving the signal-to-noise of fluorescence images. Particularly high numerical apertures can be obtained using solid-immersion approaches, where the imaged atom plane is only a few microns away from a final hemispherical lens, which enhances the numerical aperture of the objective by the index of refraction of the hemisphere~\cite{bakr2009quantum}. This has allowed reaching numerical apertures of up to 0.87~\cite{parsons2015site,cheuk2015quantum}. 

The second requirement for a microscope experiment is the preparation of a 2D sample. Compared to a 3D lattice, this avoids the issue of out-of-focus fluorescence in the presence of multiple layers, given the micron-scale depth of focus of high NA imaging systems. Multiple approaches have been used to prepare 2D samples. In an early iteration of the Harvard boson microscope, which uses an in-vacuum solid immersion lens, a magnetic trap was used to push the cloud against a repulsive barrier created using an evanescent wave trap at the lens surface~\cite{gillen2009two}. This compressed the gas so that it could be subsequently loaded into a single layer of standing wave trap created by a laser beam reflected at an angle off the surface. In other experiments, a highly anisotropic light sheet trap is used to compress the system to load a single layer of large spacing accordion lattice formed by laser beams interfering at a shallow angle. The gas is then brought deep into the 2D regime by dynamically increasing the angle between the beams to reduce the lattice spacing~\cite{parsons2015site,brown2017spin}. A third approach used by some groups is to spin-flip a single layer of atoms from a gas loaded into a one-dimensional lattice using microwave spectroscopy in a magnetic field gradient, and then subsequently remove the rest of the layers with resonant light~\cite{sherson2010single,cheuk2015quantum,omran2015microscopic,haller2015single,edge2015imaging}.

The third requirement for microscopy is obtaining fluorescence images with large signal-to-noise to enable high-fidelity reconstruction of site occupancies without significant hopping or loss of atoms during the imaging process. To do this, it is usually necessary to detect about $10^3$ photons per atom. Given the limited collection efficiency, the number of scattered photons is often more than an order of magnitude larger. Even in the very deep lattices using for pinning the atoms during imaging (a few thousand lattice recoils), it is necessary to illuminate the atoms using light in a laser cooling configuration during imaging to avoid thermal hopping and loss. The light scattered during the laser cooling is collected by the imaging system. For heavy species like $^{87}$Rb or $^{133}$Cs, polarization gradient cooling (PGC) suffices for this purpose~\cite{bakr2009quantum,sherson2010single,impertro2023unsupervised}. For lighter species, PGC is less effective, and alternative schemes have been used. For $^6$Li, Raman sideband cooling has been the method of choice~\cite{omran2015microscopic,parsons2015site,brown2017spin}, while for $^{40}$K, both Raman sideband cooling~\cite{cheuk2015quantum} and electromagnetically-induced-transparency cooling have been used~\cite{haller2015single,edge2015imaging}. Since the point-spread-function of an atom in the images is typically comparable to the lattice spacing, high-fidelity reconstruction of the occupancies is a challenging task, but various algorithms have been developed, including ones based on machine learning, and percent level infidelities in detection are standard~\cite{impertro2023unsupervised,la2023comparative}.

In the standard imaging approach with quantum gas microscopes, light-assisted collisions present a limitation on the information that can be extracted from the images. If a site is occupied by more than one atom during imaging, the atoms are rapidly ejected pairwise due to light-assisted collisions and are not detected. This leads to a detection of the atom number on a site modulo two. For single-band Fermi-Hubbard systems, the maximum occupancy on a site is two (one spin-up and one spin-down atom), and doubly occupied sites appear empty in the images. This limitation has been circumvented in later experiments using bilayer imaging schemes~\cite{preiss2015quantum,koepsell2020robust,hartke2020doublon,yan2022two,mongkolkiattichai2025quantum}. The spin-up and spin-down atoms can be separated just before imaging into different, well-separated layers, for example using a magnetic field gradient in a Stern-Gerlach protocol. The separation is chosen to be much larger than the depth of focus of the imaging system. The two layers can then be simultaneously focused and imaged onto a camera using two ``eyepiece" lenses. With this approach, the four possible states on a lattice site of a Fermi-Hubbard system may be distinguished (empty, up-spin, down-spin or a doublon consisting of an up-spin and a down-spin). 

\subsection{Realizing the Hubbard model and tuning its parameters}
The Fermi-Hubbard model is realized by loading the two-component Fermi gas into an optical lattice. An optical lattice is a periodic pattern of light created by interfering laser beams. The atoms experience the lattice as a periodic potential due to the AC Stark effect. For example, a laser beam with wavenumber $k$ propagating along the $x$ direction interfering with its retroreflection results in a standing wave with the corresponding potential $V(x)=-V_0\cos^2{kx}$, where $V_0$ is the lattice depth. Here we assume the light is red-detuned relative to the atomic resonance and the atoms are trapped at the intensity maxima. The natural energy scale associated with this potential is called the recoil energy and is given by $E_R = \hbar^2 k^2/ 2m = \hbar^2 \pi^2/ 2ma^2$, where $m$ is the mass of the atoms and $a$ is the lattice spacing. It is used as the unit for reporting lattice depths in experiments. A 2D square lattice may be realized by introducing a second retroreflected beam propagating along the $y$-axis detuned by a few tens of MHz to time-average away the interference between the two pairs of beams. According to Bloch's theorem, the solutions of the Schr\"{o}dinger equation in such a potential take the form of energy bands $E^{(n)}_\mathbf{q}$ labelled by the band index $n$ and the quasimomentum $\mathbf{q}$. A degenerate Fermi gas with chemical potential below the recoil energy can be loaded adiabatically into the lowest band of the lattice, allowing for the realization of a single-band model. 

The second quantized Hamiltonian describing the two-component Fermi gas in the lattice potential $V(\bf{r})$ is

\begin{equation}
\hat{H} = \sum_{\sigma\in\{\uparrow,\downarrow\}} \int d\mathbf{r}~\hat{\Psi}^\dagger_\sigma(\mathbf{r})\left[\frac{-\hbar^2\nabla^2}{2m}+V(\mathbf{r})\right]~\hat{\Psi}_\sigma(\mathbf{r}) + g\int d\mathbf{r} ~\hat{\Psi}^\dagger_\uparrow(\mathbf{r})\hat{\Psi}^\dagger_\downarrow(\mathbf{r})\hat{\Psi}_\downarrow(\mathbf{r})\hat{\Psi}_\uparrow(\mathbf{r}),
\end{equation}
where $\hat{\Psi}^\dagger_\sigma(\mathbf{r})$ is the field operator for creating a fermion of spin $\sigma$ at position $\mathbf{r}$ and the coupling $g$ characterizes the scattering between the spin-up and spin-down fermions. The Fermi-Hubbard model can be obtained by expanding the field operators in terms of the site operators $\hat{c}_{j,\sigma}$ used in Eq.\ref{eq:hubbard_model}
\begin{equation}
\hat{\Psi}_\sigma(\mathbf{r}) = \sum_j w_j(\mathbf{r}) \hat{c}_{j,\sigma},
\end{equation}
where the weights are given by the Wannier wavefunctions $w_j(\mathbf{r})$~\cite{jaksch1998cold}. These are superpositions of Bloch states maximally localized on lattice site $j$. The first term of the Hamiltonian then reduces to a tunneling Hamiltonian $-\sum_{j,l>j,\sigma} t_{j,l} \hat{c}^\dagger_{j,\sigma}\hat{c}_{l,\sigma} +$ h.c., where the tunneling matrix element between sites $j$ and $l$ is given by
\begin{equation}
t_{j,l} = -\int d\mathbf{r}~w^{*}_j(\mathbf{r})\left[-\hbar^2\nabla^2/2m + V(\mathbf{r})\right] w_l(\mathbf{r}).
\end{equation}
For the 2D square lattice potential discussed above, tunnelings beyond nearest-neighbor become negligible for lattice depths above $\sim5~E_R$, reducing the kinetic term to that appearing in Eq.~\ref{eq:hubbard_model}. A typical value for the nearest neighbor tunneling rate for $^6$Li is a few hundred Hz.

At low temperatures, scattering between atoms is in the $l=0$ partial wave and is characterized by a scattering length $a_s$. The complicated microscopic potential can be replaced by a $\delta$-function interaction potential that reproduces the scattering length. The relationship between the coupling $g$, which characterizes scattering of atoms with opposite spin states, and the scattering length is given by $g = 4\pi \hbar^2 a_s/m$. Using the expansion of the field operators in the second term of the continuum Hamiltonian leads to the tight-binding model, where similarly to the kinetic term, off-site interactions are negligible in a sufficiently deep lattice, and only the on-site term is retained. This leads to the Hubbard interaction term in Eq.~\ref{eq:hubbard_model} with $U = g \int d\mathbf{r}~|w_0(\mathbf{r})|^4$.

In principle, the exponential dependence of the tunneling $t$ on the lattice depth allows tuning the ratio $U/t$ over orders of many magnitude. In practice, the minimum tunneling used in experiments is limited by the need to adiabatically transform the state of the system within the timescale set by heating or decoherence in the experiment. Additionally, the scattering length $a_s$, and hence $U$, can be tuned using a bias magnetic field by working close to a Feshbach resonance \cite{chin2010feshbach}. The interaction $U$ is constrained on the high end to be a fraction of the bandgap for the validity of a single-band description.

A Feshbach resonance is a resonance between a continuum state of two colliding atoms in one interatomic potential and a weakly bound state in another. The potentials correspond to the valence electrons of the two atoms having a different total spin, and hence different magnetic moments. This allows tuning the relative energy of the states and bringing them into resonance where the scattering length diverges. For example, such a resonance exists in $^6$Li atoms in the lowest two hyperfine states at 832~G~\cite{zurn2013precise}. Below the resonance, the scattering length between the atoms is positive and a molecular state emerges. The molecular state can be populated through three-body collisions very close to the resonance, but further away, the atoms remain in the metastable upper branch over the duration of the experiment (hundreds of milliseconds), which permits the realization of a Fermi-Hubbard model with tunable repulsive interactions. The typical interaction strength $U/h$ for $^6$Li is on the order of a few kHz.

\subsection{Engineering the trapping potential}
\label{sec:engineering_potential}

\begin{figure}[t]

%\sidecaption
% For example, with the graphicx style use
\centering
\includegraphics[width=10cm]{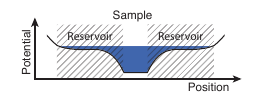}
%
% If no graphics program available, insert a blank space i.e. use
%\picplace{5cm}{2cm} % Give the correct figure height and width in cm
%
\caption{The overall confinement in the lattice can be engineered with a digital micromirror device to cool the system through entropy redistribution. Entropy is transferred from the gapped Mott insulating sample region to the surrounding gapless metallic reservoir. Adapted from~\cite{mazurenko2017cold}}

\label{fig:dmd}       % Give a unique label
\end{figure}

Unlike real materials, where the electron density is usually spatially homogeneous when averaged over a few unit cells, the density in an atomic gas varies spatially due to the trapping potential. When Gaussian laser beams are used to create the lattices, the trapping potential is approximately harmonic near the center of the trap. In most cases, the local density approximation (LDA) is an excellent approximation for calculating properties of the gas. If the trapping potential varies slowly, a small patch of the gas can be treated as a homogeneous system with local chemical potential $\mu(\mathbf{r})$ in the grand canonical ensemble. An extra term is then appended to the Hubbard model of Eq.~\ref{eq:hubbard_model} of the form $-\sum_i \mu_{i}(\hat{n}_{i\uparrow}+\hat{n}_{i\downarrow})$. For example, for a harmonic potential, $\mu_i = \mu_0 - \frac{1}{2} m \omega^2 a^2 r_i^2$, where $\omega$ is the harmonic trap's frequency and $\mu_0$ is a global chemical potential that fixes the total atom number in the trap.

The naturally present harmonic trapping potential is useful in two ways. First, the spatially varying density it produces gives the experimentalist access to a cut through the phase diagram along the doping axis within the same cloud. In this way, correlations can be obtained vs. doping without having to change the total atom number in the trap. For example, the local chemical potential may be appropriate for forming a Mott insulator in the center of the trap, but the lower chemical potential towards the edge of the cloud would correspond to a lower filling that gives rise to a metallic state. A second advantage of the trap is in cooling the gas through a mechanism known as entropy redistribution. In thermal equilibrium, the temperature is fixed throughout the cloud, but the local entropy per particle varies. At the same temperature, a region with a gapped phase like the Mott insulator will not be able to accommodate as many thermal excitations as a non-gapped phase like a metal. In other words, the Mott phase has a lower specific heat capacity than the metal. Therefore, when adiabatically turning on the lattice potential, entropy is naturally expelled from the Mott insulating region to the metal. This reduction of entropy in the Mott insulator region results in a lower global temperature than if the system were in a homogeneous Mott insulator~\cite{bernier2009cooling}. 

It is clear from the discussion above that precise control over the trapping potential is highly desirable. This can be obtained by projecting light patterns tailored with spatial light modulators through the objective onto the cloud. Microscope experiments have typically used digital micromirror devices (DMDs) for this purpose~\cite{liang2010high,gauthier2016direct,zupancic2016ultra}. With this approach, the Harvard group has engineered the trapping potential to further enhance entropy distribution (Fig.~\ref{fig:dmd})~\cite{mazurenko2017cold,chiu2018quantum}. In their experiment, a central homogeneous trapping region was produced, surrounded by a low-density metallic reservoir. The chemical potential in the homogeneous region could be varied to study different regimes of the Fermi-Hubbard model. Besides improving the entropy redistribution, this approach is helpful in another way. At low temperatures, correlation lengths (e.g., of antiferromagnetic order) can become large and eventually limited by the density variations in a harmonic trap. The uniform central trapping region addresses this problem.

Finally, control over the potential landscape of the atoms using DMDs can be used to prepare initial density profiles for non-equilibrium dynamics experiments as discussed in Sec.~\ref{sec:transport}.

\section{Fermi-Hubbard systems at half-filling}
At half-filling and strong interactions, the Fermi-Hubbard model ground state is an antiferromagnetic Mott insulator. The interaction term dominates over the tunneling term in the Hamiltonian, and suppresses costly density fluctuations, leading to exactly one atom per site in the infinite $U/t$ limit. Since the half-filled regime is well-understood theoretically, it was the focus of much of the early experimental work with cold atoms, even before fermionic quantum gas microscopes were introduced. In 2008, the ETH and MPQ groups reported the first observations of fermionic Mott insulators~\cite{jordens2008mott,schneider2008metallic}. These observations were based on the suppression of doubly occupied sites (measured spectroscopically through their interaction energy shifts), the appearance of a gapped mode in the excitation spectrum obtained by modulating the lattice depth, and the emergence of an incompressible phase as identified in the response of the cloud size to changing the harmonic confinement. Subsequent work from the ETH and Rice groups observed short-range antiferromagnetic correlations~\cite{greif2013short,hart2015observation}. The Rice group used spin-sensitive Bragg scattering of light to measure the correlations in the 3D Hubbard model, demonstrating temperatures 1.4 times the critical temperature for the phase transition to the antiferromagnet. With the advent of fermionic quantum gas microscopes, the study of the Fermi-Hubbard model with cold atoms made rapid advances as discussed in the next sections.

\subsection{Mott insulators}

\begin{figure}[t]

%\sidecaption
% For example, with the graphicx style use
\centering
\includegraphics[width=\textwidth]{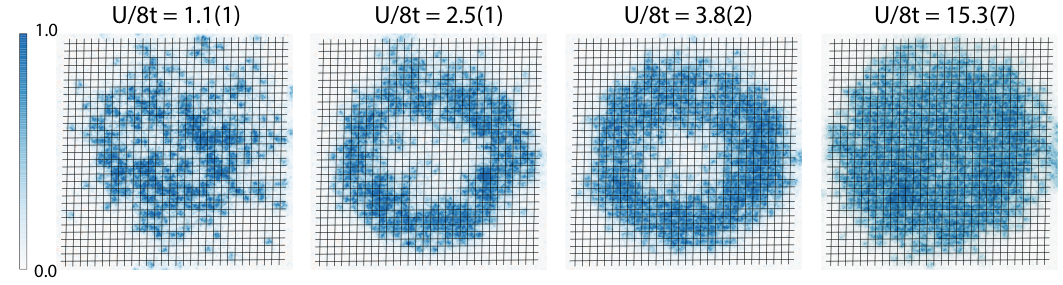}
%
% If no graphics program available, insert a blank space i.e. use
%\picplace{5cm}{2cm} % Give the correct figure height and width in cm
%
\caption{Transition from a metal to a Mott insulator by increasing the interaction strength $U/t$ using a Feshbach resonance. The metal in the leftmost image exhibits strong density fluctuations, which are suppressed in the Mott insulator in the rightmost image. The middle two images illustrate a central band insulator, which appears empty due to parity-projected imaging, surrounded by a Mott insulating ring. The gases in these images are not at the same entropy, so the evolution shown cannot be understood to be the result of purely increasing $U/t$. Adapted from~\cite{greif2016site}.}

\label{fig:mottinsulator}       % Give a unique label
\end{figure}

Fermionic Mott insulators were imaged with single-site resolution by the Harvard group with $^6$Li and the MIT group with $^{40}$K~\cite{greif2016site,cheuk2016observationmott}. Fig.~\ref{fig:mottinsulator} shows site-resolved images illustrating the metal to Mott insulator transition at constant atom number as the interaction strength $U/t$ is increased using a Feshbach resonance~\cite{greif2016site}. For the weakest interaction, the metallic state manifests through a large variance of the site occupation. For the strongest interactions, a large half-filled region is observed where the variance in the occupation is significantly suppressed to about 2\% at the center of the cloud. At intermediate interactions, a clear shell structure is observed, where the center is a band insulator that appears empty due to light-assisted collisions, and is surrounded by a Mott insulating ring.

\begin{figure}[t]

%\sidecaption
% For example, with the graphicx style use

\includegraphics[width=\textwidth]{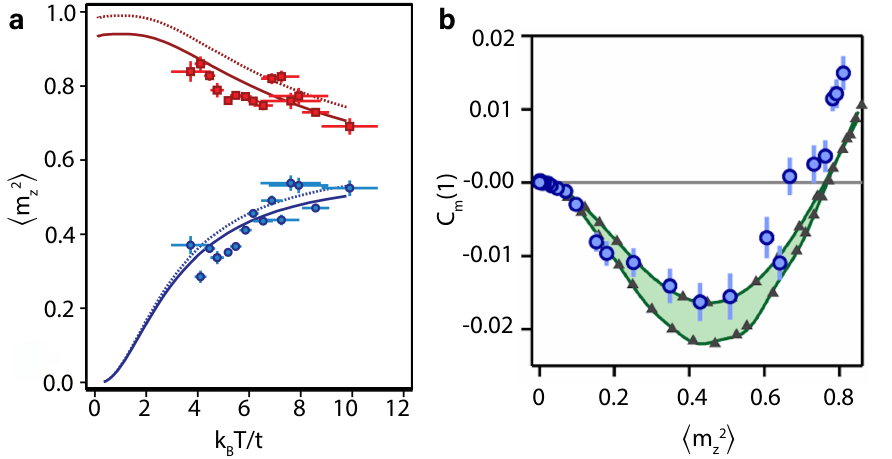}
%
% If no graphics program available, insert a blank space i.e. use
%\picplace{5cm}{2cm} % Give the correct figure height and width in cm
%
\caption{a. Local moment vs. temperature at $U/t\approx21$. The red curve (squares) corresponds to half-filling, while the blue curve (circles) corresponds to $\mu/U = -0.25$. b. Density correlations vs. local moment. Adapted from ~\cite{cheuk2016observationmott,cheuk2016observationcorrelations}.}
\label{fig:localmoment}       % Give a unique label
\end{figure}

The parity-projected density detected on a site in these experiments is equivalent to a quantity known as the local moment. Its expectation value is given by $\langle \hat{m}_z^2\rangle = \langle(\hat{n}_\uparrow - \hat{n}_\downarrow)^2\rangle = \langle \hat{n}_\uparrow + \hat{n}_\downarrow - 2\hat{n}_\uparrow \hat{n}_\downarrow\rangle$, and is sometimes known as the ``singles" density. The dependence of the local moment on temperature in a strongly-interacting, half-filled system is shown by the red points in Fig.~\ref{fig:localmoment}a~\cite{cheuk2016observationmott}. As the system enters the Mott insulating state at low temperatures, the local moment approaches unity. At the high temperatures for this data ($T/t>4$), both high-temperature series expansion (HTSE) and numerical linked cluster expansion (NLCE) numerics work quite well and are in agreement with the experimental results.

Another interesting feature of the Mott insulator can be discerned by looking at the connected nearest-neighbor correlation function of the local moments given by 
\begin{equation}
C_m = \langle \hat{m}^2_{z,i} \hat{m}^2_{z,j}\rangle - \langle \hat{m}^2_{z,i}\rangle \langle \hat{m}^2_{z,j}\rangle,
\end{equation}
where $i$ and $j$ are nearest neighbor sites. This is shown in Fig.~\ref{fig:localmoment}b vs. the local moment for $U/t\sim 7$~\cite{cheuk2016observationcorrelations}. For low moments, corresponding to a metallic system, a negative correlation is observed. This is due to a combination of two effects that lead to fermion antibunching. For fermions of the same spin, Pauli exclusion leads to a ``Pauli hole". For fermions of opposite spins, there is a correlation hole due to the strong repulsive interactions. As the local moment increases, the system enters the Mott insulating regime. Although most of the sites are occupied by one atom, the finite $U/t$ leads to quantum fluctuations, where an atom can briefly hop onto a neighboring one, introducing virtual ``doublon-hole" pairs that can be captured in the snapshots. Due to parity imaging, these appear as two neighboring holes. The clustering of holes in the Mott insulator snapshots means that there is a corresponding clustering of singles, which explains the positive moment correlations observed as the local moment approaches unity.

\subsection{Antiferromagnets}
\label{sec:antiferromagnets}

\begin{figure}[t]

%\sidecaption
% For example, with the graphicx style use
\centering
\includegraphics[width=\textwidth]{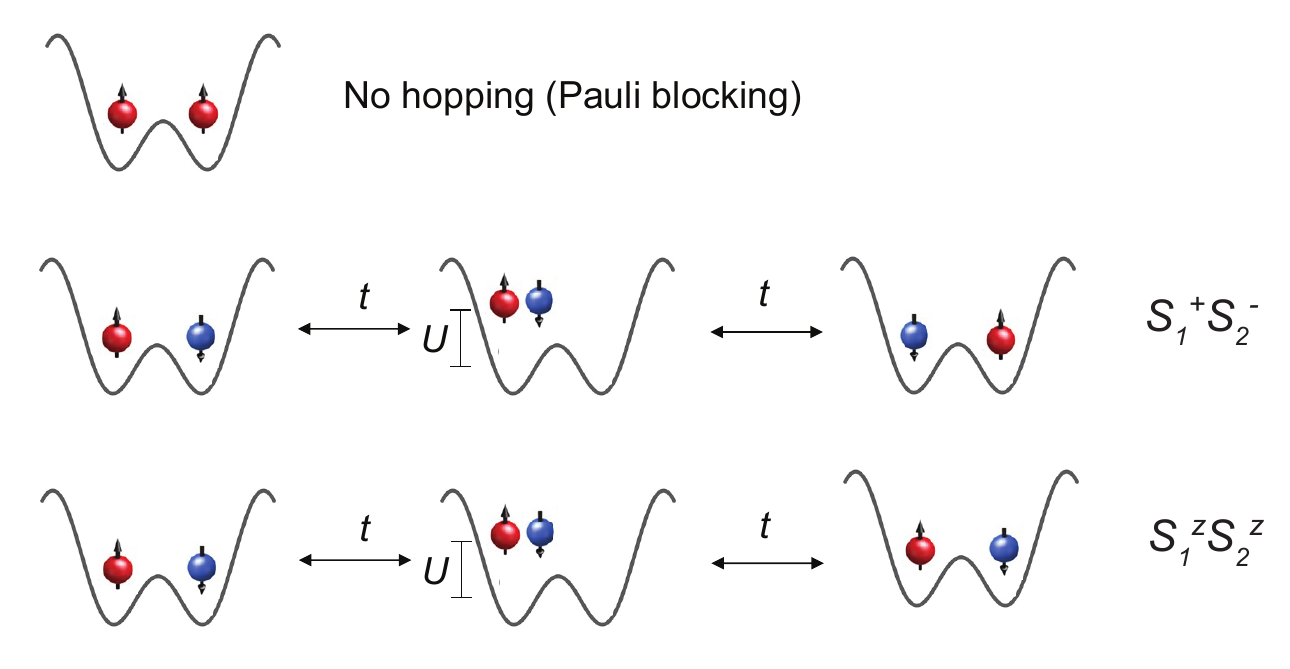}
%
% If no graphics program available, insert a blank space i.e. use
%\picplace{5cm}{2cm} % Give the correct figure height and width in cm
%
\caption{Cartoon derivation of the terms in the effective Heisenberg Hamiltonian at half-filling and large $U/t$. If the spins in the double-well are the same, there is no energy shift since Pauli blocking suppresses hopping. For different spins, second-order processes lead to spin-exchange or Ising terms depending on whether the spins swap or not. Each of these processes can occur in two different ways.}
\label{fig:superexchange}       % Give a unique label
\end{figure}

The Fermi-Hubbard Hamiltonian in Eq.~\ref{eq:hubbard_model} only has an on-site interaction. Nevertheless, strong correlations between spins on different sites can develop in the Mott insulator at low temperatures. These correlations result from a second order process known as superexchange, which favors anti-alignment of neighboring spins. This can already be understood at the level of two atoms in a double well in the large $U/t$ limit (Fig.~\ref{fig:superexchange}). Ignoring tunneling to lowest order, the state of the double-well system has one atom one each site to avoid double-occupancies which are energetically costly. If the spins of the atoms are aligned, the atoms cannot tunnel anyway due to Pauli exclusion. However, if the atoms have anti-aligned spins, a second-order process can occur where an atom hops with matrix element $t$ onto the neighboring site, and then hops back, or alternatively the two spins can exchange positions. The intermediate state is energetically detuned by $U$ from the initial and final states. Thus, according to second-order perturbation theory, the energy of this configuration with anti-aligned spins is lowered relative to the configuration with aligned spins by an amount $\propto t^2/U$. More formally, a low-energy effective Hamiltonian known as the $t-J-3s$ model can be obtained from the Hubbard model in the limit of large $U/t$ by using a Schrieffer-Wolff transformation to eliminate double occupancies~\cite{kale2022schrieffer}. At half-filling, this Hamiltonian reduces to the Heisenberg model whose Hamiltonian is

\begin{equation}
\hat{H} = J \sum_{\langle i,j\rangle} \mathbf{\hat{S}}_i \cdot \mathbf{\hat{S}}_j,
\end{equation}
where the exchange coupling is $J=4t^2/U$. The vector spin operator on site $i$ is given by $\hat{\mathbf{S}}=\sum_{a,b}\hat{c}^\dagger_{i,a} \boldsymbol\sigma_{a,b}\hat{c}_{i,b}$, with $\boldsymbol\sigma$ being the vector of Pauli matrices. 

The Hubbard model possesses an SU(2) spin symmetry. According to the Mermin-Wagner theorem, the two-dimensional Hubbard model with such a continuous symmetry cannot have long-range spin order at finite temperature. The two-point spin correlations decay exponentially at any finite temperature, with the correlation length diverging exponentially with the inverse temperature. This is in contrast to the 3D Hubbard model, where there is a finite-temperature phase transition to an antiferromagnet which spontaneously breaks the SU(2) spin symmetry. This phase transition has been recently observed with cold atoms~\cite{shao2024antiferromagnetic}.

\begin{figure}[t]

%\sidecaption
% For example, with the graphicx style use
\centering
\includegraphics[width=\textwidth]{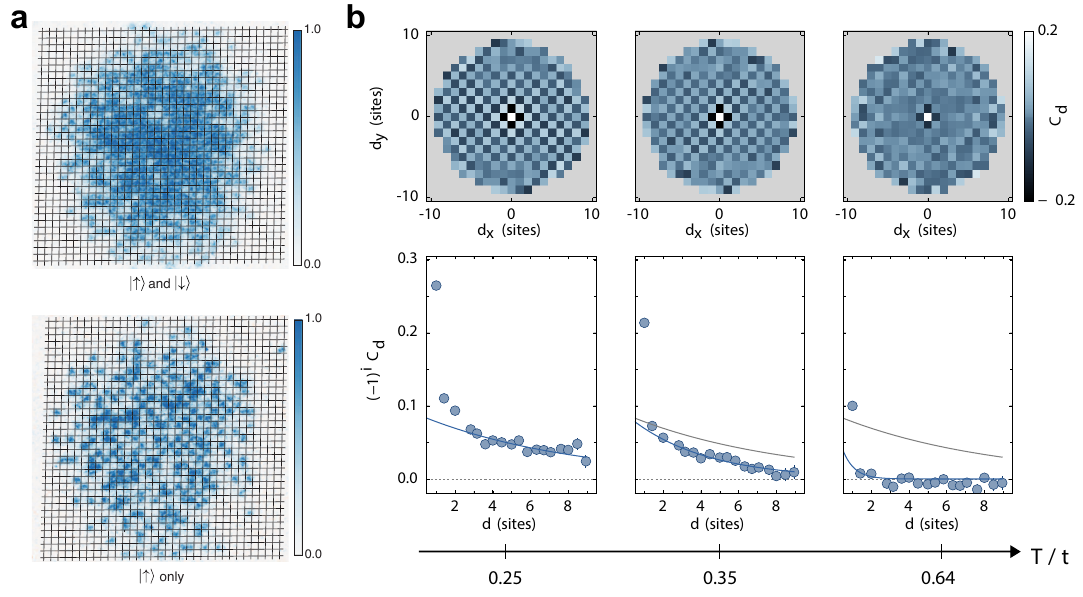}
%
% If no graphics program available, insert a blank space i.e. use
%\picplace{5cm}{2cm} % Give the correct figure height and width in cm
%
\caption{a. Images of Mott insulator before (top) and after (bottom) removing one of the spin states with a resonant pulse of light. b. Evolution of the two-point spin correlation function with temperature $T/t$. Adapted from~\cite{mazurenko2017cold,parsons2016site}.}
\label{fig:spin_removal}       % Give a unique label
\end{figure}

The two-point spin correlation function between sites $i$ and $j$ is given by
\begin{equation}
\label{eq:spin_correlation}
C_s= 4 \left(\langle \hat{S}^z_i \hat{S}^z_j\rangle - \langle \hat{S}^z_i\rangle\langle\hat{S}^z_j\rangle\right).
\end{equation}
A displacement-dependent spin correlation function $C_s(\mathbf{d})$ can be defined by averaging over all sites within a given region that are separated by site displacement $\mathbf{d}$. This quantity can be extracted using quantum gas microscopy even without bilayer imaging~\cite{parsons2015site,cheuk2016observationcorrelations}. The key idea is to use a short resonant pulse of light to selectively remove one of the spin components, along with any double occupancies due to the ensuing light-assisted collisions. Images of a cloud with a large Mott insulating region are shown in Fig.~\ref{fig:spin_removal}a, before and after the spin removal pulse. One can easily discern by inspection the appearance of checkerboard type correlations in the image after spin removal, indicating antiferromagnetic order. The correlations are not expected to be perfect, not only due to the fact that the system is at finite temperature, but also because of quantum fluctuations. Even at zero temperature, the N\'{e}el vector can point in any direction on the Bloch sphere, and strong checkerboard order is only detected when it is aligned with the measurement basis. It can be shown that $C_s = 2(C_\uparrow + C_\downarrow)-C_m$, where $C_\sigma$ is the particle correlation function obtained after a spin removal pulse that keeps only atoms of spin $\sigma$ on singly occupied sites. All three quantities on the right-hand side of the equation can be separately measured. Fig.~\ref{fig:spin_removal}b shows $C_s(\mathbf{d}$) obtained this way by the Harvard group vs. temperature, evaluated in a Mott insulating region~\cite{mazurenko2017cold}. The correlations can be quantitatively compared to quantum Monte Carlo numerics with the temperature as a fit parameter. Typical temperatures obtained in such experiments are in the range $T/t = 0.25$ to $0.5$.

So far, we have focused on spin-balanced antiferromagnets. The Princeton group has performed experiments exploring spin-imbalanced half-filled systems which break the SU(2) spin symmetry of the Hubbard model~\cite{brown2017spin}. A spin-imbalanced gas can be prepared by evaporating the gas in the presence of a magnetic field gradient. A difference in the magnetic moments of the two spin components leads to different evaporation rates. In such a gas, the absence of spin-changing collisions ensures that the spin populations are conserved after the end of evaporation. The spin polarization can be taken into account in the Hubbard Hamiltonian by introducing an effective Zeeman field $h$ that couples to $\hat{S}^z$. 

\begin{figure}[t]
    \centering
    \includegraphics[width=\textwidth]{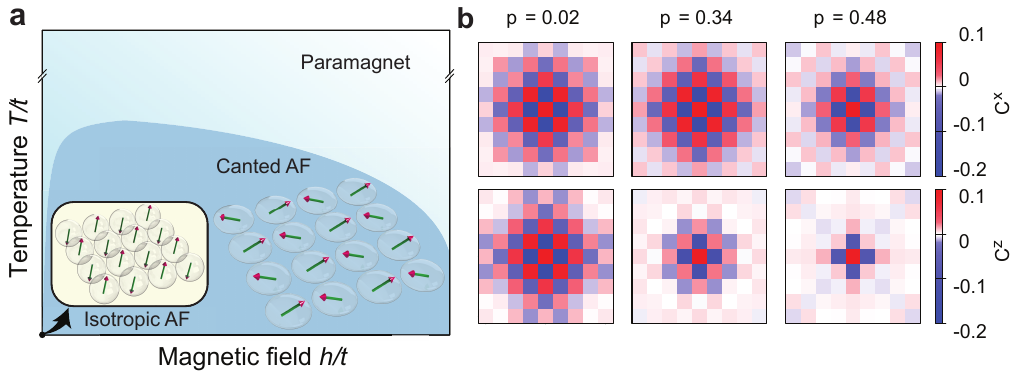}
    \caption{a. Phase diagram of the Heisenberg model in the presence of an effective Zeeman field. b. Two-point spin correlation function in the direction orthogonal to (top row) and along (bottom row) the effective field vs. spin polarization in the Mott insulator. The correlations in the plane orthogonal to the field become dominant, a precuror to long-range order in the canted antiferromagnet. Adapted from~\cite{brown2017spin}.}
    \label{fig:cantedAF}
\end{figure}

A cartoon phase diagram for a Mott insulator described by a Heisenberg model in the presence of an effective field is shown in Fig.~\ref{fig:cantedAF}a~\cite{brown2017spin}. For non-zero $h$, the SU(2) spin symmetry is reduced to a U(1) symmetry. The 2D system can then undergo a Berezinskii–Kosterlitz–Thouless (BKT) phase transition to an ordered phase below a finite critical temperature where spin vortices and anti-vortices bind together. The ordered phase is a canted antiferromagnet with long-range spin order in the plane orthogonal to the effective field. In the experiment, the cloud was above this critical temperature. Nevertheless, precursor correlations to this phase were observed, seen in Fig.~\ref{fig:cantedAF}b vs. polarization. While the spin correlations along the field and in the plane are almost symmetric near zero polarization (as expected for an SU(2) symmetric system), the in-plane correlations evolve to become stronger than the correlations along the field at large polarizations.

\section{Doped Fermi-Hubbard systems}

\begin{figure}[t]
    \centering
    \includegraphics[width=\textwidth]{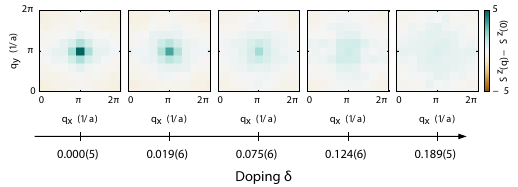}
    \caption{Measured spin structure factor vs. doping in the square lattice model. Antiferromagnetic correlations vanish quickly with doping in 2D. Adapted from~\cite{mazurenko2017cold}.}
    \label{fig:spin_structure_factor}
\end{figure}

In the Mott insulator, the charge degree of freedom is frozen.\footnote{Here we use the word ``charge" loosely to refer to density because of the analogy with electronic systems, despite the neutrality of cold atom systems.} However, most of the interesting physics of the Fermi-Hubbard model results from the interplay of the charge and spin degrees of freedom upon doping the antiferromagnetic insulator. 

On the square lattice, doping suppresses antiferromagnetism. This can be seen most clearly in the spin structure factor, corresponding to the Fourier transform of the spin correlation function. Fig.~\ref{fig:spin_structure_factor} shows the experimentally measured evolution of the structure factor with doping. In the half-filled system, there is a narrow peak at momentum $\mathbf{q}=(\pi,\pi)/a$. The peak is quickly suppressed with hole doping, and almost completely vanishes by $\delta \sim 15\%$. Qualitatively, this is the same behavior seen in the phase diagram in Fig.~\ref{fig:cuprates}, although in the cuprates, the antiferromagnetism vanishes at smaller doping, indicating that the Hubbard model needs to refined with more terms. In the cuprates, the static spin structure factor can be measured using elastic neutron scattering. However, as we will see next, in comparison to solid state probes, cold atom microscopy snapshots provide much more detailed information about doped Hubbard systems beyond the static spin structure factor.

\subsection{Doped Hubbard chains}
To better appreciate the richness of the doped 2D Fermi-Hubbard model, it is worth starting with the 1D case. Interestingly, the interplay between charge and spin is absent in this case! In fact, the spin and charge degrees of freedom live in independent sectors, a phenomenon known a spin-charge separation~\cite{giamarchi2003quantum}. While this phenomenon is well-understood theoretically, experiments from the MPQ group enabled the experimentalists to observe it in a very direct way~\cite{hilker2017revealing,vijayan2020time}.

\begin{figure}[t]
    \centering
    \includegraphics[width=\textwidth]{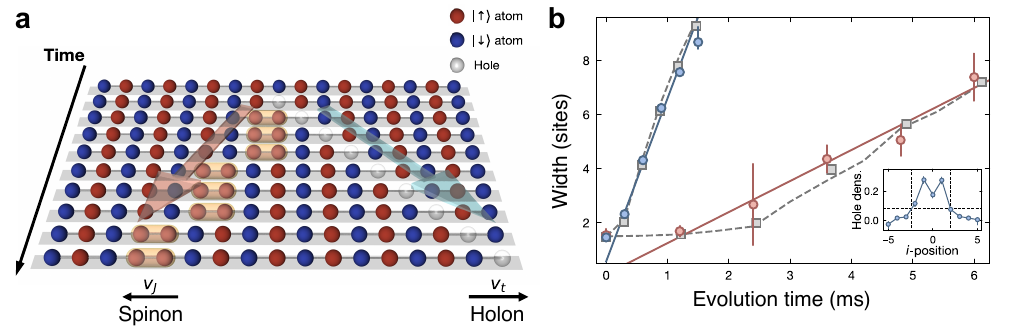}
    \caption{a. After injection of a hole into a Fermi-Hubbard chain, the hole breaks apart into a spinon and a holon which carry fractional quantum numbers. This is called spin-charge separation. b. The holon moves with a speed set by $t$ (blue line), while the spinon moves more slowly with a speed set by $J$ (red line). Adapted from~\cite{vijayan2020time}.}
    \label{fig:spincharge_separation}
\end{figure}

In one experiment, a hole excitation was introduced in a Hubbard chain by using a focused resonant pulse of light to remove an atom. Fig.~\ref{fig:spincharge_separation}a illustrates the dynamics of the hole in an antiferromagnetic chain\footnote{The pictures used in this section use Ising spins and ignore quantum fluctuations. They provide an intuitive understanding of the physics, despite the effective model at half-filling being a quantum Heisenberg model.}~\cite{vijayan2020time}. When the hole is introduced, it is clear that the spins surrounding it should be positively correlated with each other (aligned). However, after a few hops, the hole fragments into two excitations that ``forget" about each other. One is a holon, a charged particle surrounded by anti-aligned spins. The other is a pair of aligned spins (a domain wall), known as a spinon. Both are long-lived quasiparticles that propagate through the chain. The injected hole has charge -1 and spin 1/2 as its quantum numbers. However, it is not a stable quasiparticle in the many-body system. It fractionalizes, with the charge carried by the holon and the spin is carried by the spinon. It is clear from this simple picture that the two excitations have different speeds. The holon dynamics are set by the tunneling $t$, while the spinon dynamics are set by the superexchange $J$. All these expectations can be made quantitative within the framework of Luttinger liquid theory. Fig.~\ref{fig:spincharge_separation}b shows the velocities of the two excitations measured in the experiment, which are very different indeed~\cite{vijayan2020time}. 

\begin{figure}[t]
    \centering
    \includegraphics[width=.7\textwidth]{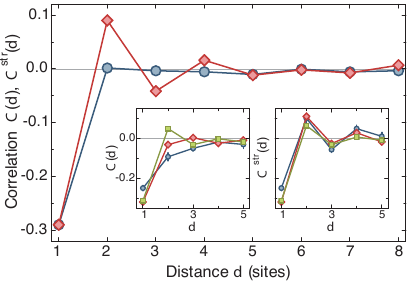}
    \caption{In a Fermi-Hubbard chain, the amplitude of the two-point spin correlation is strongly suppressed with doping (blue line, circles). However, the ``hidden" spin correlations can be revealed by considering a string correlator (red line, diamonds). Adapted from~\cite{hilker2017revealing}.}
    \label{fig:string_correlators}
\end{figure}

In fact, spin-charge separation can be observed in doped chains in thermal equilibrium, without the experimentalists introducing any excitations. For a hole-doped system, it is energetically favorable for the mobile dopant to be surrounded by anti-aligned spins as we have seen. In other words, the dopant acts as a domain boundary, flipping the orientation of the spins that come after it relative to the situation with the dopant absent. If one studies the system with the two-point correlation function defined in Eq.~\ref{eq:spin_correlation}, the spin correlations in the chain would quickly decay with doping due the shifts of the spin pattern introduced by each hole. However, a non-local ``string" correlator can be defined, which introduces a flip in the sign of the spin each time one crosses a hole in a snapshot. Compared to the normal two-point correlator, this string correlator reveals the hidden spin correlations in the doped system, as seen in the data in Fig. ~\ref{fig:string_correlators}~\cite{hilker2017revealing}.

\subsection{Geometric strings and magnetic polarons in 2D}
\label{sec:magnetic_polarons}

\begin{figure}[t]
    \centering
    \includegraphics[width=\textwidth]{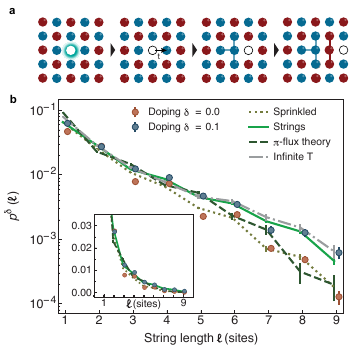}
    \caption{a. A hole moving in a 2D antiferromagnet leaves in its wake a string of bonds of the incorrect sign. The string attaches the holon to a spinon that remains behind and moves more slowly. b. Probability of observing a string of a particular length in quantum gas microscope snapshots. For the doped 2D system (blue points), the probability of observing long strings is consistent with the expectation from the geometric string theory (green line) and is higher than the probability of observing strings in simulations where holes are randomly sprinkled with the correct density onto an antiferromagnet. Adapted from~\cite{ji2021coupling,chiu2019string}.}
    \label{fig:geometric_string}
\end{figure}

Moving from 1D to 2D, the motion of a dopant in an antiferromagnetic environment displays qualitatively different physics~\cite{Schmitt1988,Shraiman1988,Sachdev1989,Kane1989,Dagotto1989}. As shown in Fig.~\ref{fig:geometric_string}a, a hole moving in a square antiferromagnet leaves behind a spinon as before, but also a string of bonds of the wrong sign in its wake. The energy cost associated with this string grows linearly with the distance $l$ covered by the holon, leading to a holon-spinon confinement mechanism. This potential energy, proportional to $J\times l$, competes with the tunneling $t$ that favors delocalization of the holon, leading to an emergent length scale in the problem. Moreover, the composite object, formed by the holon and spinon connected by a string, has a rovibrational excitation spectrum that has been studied theoretically~\cite{grusdt2018parton}. 

The Harvard group found evidence of these ``geometric strings" in quantum gas microscope snapshots~\cite{chiu2019string}. They developed an algorithm to identify strings based on the deviation of the snapshots of the doped 2D system from perfect checkerboard order, an approximation of the state of the system at half-filling. Fig.~\ref{fig:geometric_string}b shows the probability of finding a string of a particular length with their algorithm for a system with $10\%$ doping. They compare their data with simulations that obtain string length probabilities either based on (i) simply randomly inserting the correct number of holes into snaphshots obtained at half-filling, or (ii) randomly propagating holes sprinkled this way to generate strings whose lengths are drawn from a distribution predicted by the geometric string theory. The latter simulations agree much better with the data, lending credence to the string picture.

Magnetic polarons provide an alternative picture to understand the physics of the lightly doped 2D antiferromagnets. In the strong coupling limit, the superexchange coupling $J$ is small compared to the tunneling $t$. Therefore, a Born-Oppenheimer approach may be utilized. The spinon moves slowly with rate $J$, setting the bandwidth of the composite quasiparticle, while the holon at the other end of the string moves fast with rate $t$. The rapidly fluctuating string results in a local disturbance of the antiferromagnetic ordering around the mobile dopant and the emergence of a magnetic polaron. A polaron is the name given to a quasiparticle in a many-body system resulting from the coherent dressing of an impurity by its environment.\footnote{Polarons were introduced by Landau and Pekar in the context of the dressing of electrons propagating through a crystal by virtual phonons~\cite{landau1933electron,pekar1946local}. The name comes from the fact that the electrons induces a local polarization or deformation of the lattice that propagates with it. The polaron concept has since been extended to many other types of impurities and background environments. In fact, we will encounter two other types of polarons in these lectures.} In this case, the dopant is dressed by spin excitations of the antiferromagnet (magnons).

\begin{figure}[t]
    \centering
    \includegraphics[width=0.8\textwidth]{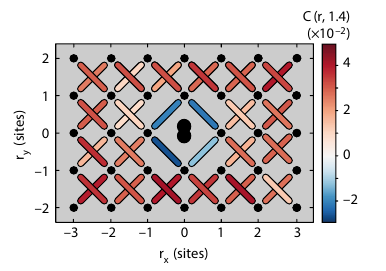}
    \caption{Measured next-nearest (diagonal) spin correlation function in the vicinity of a doublon dopant. Away from the dopant, the sign of the correlations is positive, as expected in an antiferromagnet. On the bonds around the doublon, the sign switches to negative, consistent with the predictions of a theory based on fluctuating geometric strings. Adapted from~\cite{koepsell2019imaging}.}
    \label{fig:magnetic_polaron}
\end{figure}

The MPQ group directly imaged correlations associated with these magnetic polarons~\cite{koepsell2019imaging}. In their experiment, the dopants are doublons rather than holes, but this does not affect the physics since the square lattice Fermi-Hubbard model has a particle-hole symmetry. They considered a three-point doublon-spin-spin correlation function, where a doublon is identified in the snapshot and two-point spin correlations are evaluated at different displacements in its vicinity. The next-nearest-neighbor (diagonal) spin correlation is particularly interesting and is shown in Fig.~\ref{fig:magnetic_polaron}. Far away from the doublon, this correlator is positive, as expected in a Heisenberg antiferromagnet. On the other hand, the sign of the correlations switches to negative in the immediate vicinity of the doublon, in agreement with the expectation from the string picture in Fig.~\ref{fig:geometric_string}a. The nearest-neighbor spin correlator around the dopant also gets modified, but does not exhibit a sign change. Based on these correlations, the size of the polaron is on the order of two sites. Importantly, the local disturbance of the antiferromagnetic correlations observed in the vicinity of the doublon is strongly suppressed when the doublon is pinned, emphasizing the role of the dopant mobility in the formation of the polaron. 

\begin{figure}[t]
    \centering
    \includegraphics[width=0.8\textwidth]{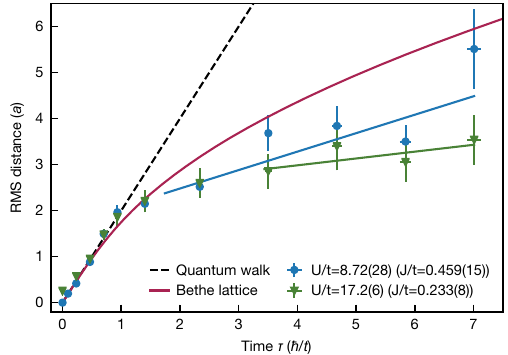}
    \caption{RMS distance characterizing the spreading dynamics of a hole injected into a 2D antiferromagnet. Initially, the hole is observed to spread ballistically with a rate set by $t$. Subsequently, the hole is dressed by magnons in the antiferromagnet to form a magnetic polaron. The polaron moves more slowly at a rate set by $J$. Adapted from~\cite{ji2021coupling}.}
    \label{fig:string_dynamics}
\end{figure}

Signatures of the magnetic polaron can also be observed in the dynamics of holes introduced into 2D antiferromagnets~\cite{ji2021coupling}. It is interesting to contrast this dynamics with the free motion of holons with a rate set by $t$ in 1D chains. Fig.~\ref{fig:string_dynamics} shows data from the Harvard group, where they observed an initial ballistic motion of the hole. However, eventually the hole ``realizes" that it is moving in antiferromagnetic environment. The dynamical formation of the polaron results in a slowdown of its motion, with further spreading of the hole determined by $J$, which sets the bandwidth of the magnetic polaron. 

As the doping of the 2D antiferromagnet is increased, polaron-polaron interactions are expected to play a more important role. Further studies with microscopes have explored the evolution of several multi-point correlators with doping, identifying a crossover from polaronic correlations to Fermi-liquid correlations at around $30\%$ doping~\cite{koepsell2020microscopic}. 

\subsection{Magnon-Fermi polarons}
Having studied how magnons in an antiferromagnet dress hole dopants, it is natural to ask how dopants dress magnon excitations in an antiferromagnet. In the doped cuprates, this question has been explored with inelastic neutron and X-ray scattering, which measure the dynamical spin structure factor~\cite{le_tacon_intense_2011,dean_persistence_2013}. The Princeton group has recently studied the dressing of magnonic impurities with dopants in the simpler setting of a spin-polarized background, observing a new type of quasiparticle, the magnon-Fermi polaron~\cite{prichard2025observation,schirotzek2009observation}. 

Fermi polarons have been studied extensively in continuum atomic Fermi gases~\cite{parish2025fermi}. A spin impurity in a polarized Fermi gas is dressed by particle-hole excitations of the Fermi sea. A simple variational wavefunction introduced by Chevy has been found to accurately predict the properties of the Fermi polaron~\cite{chevy2006universal,schirotzek2009observation},

\begin{equation}
|\Psi\rangle = \phi_0 |\mathbf{0}\rangle_\downarrow|FS\rangle_\uparrow+\sum_{|\mathbf{q}|<k_F<|\mathbf{k}|}\phi_{\mathbf{kq}} c^\dagger_{\mathbf{k}\uparrow}c_{\mathbf{q}\uparrow}|\mathbf{k-q}\rangle_\downarrow|FS\rangle_\uparrow.
\end{equation}
The first part of the superposition describes a delocalized spin-down impurity in the spin-up Fermi sea, and the second describes scattering of the impurity off the Fermi sea, resulting in a particle-hole excitation. A key difference in the lattice system is that one can start from a spin-polarized band insulator rather than a metal. Whereas in the continuum system the bare impurity is an atom, a spin-flip in an insulator is a collective excitation (a magnon). Doping the insulator with holes leads to magnon-hole scattering and the formation of the magnon-Fermi polaron.

\begin{figure}[t]
    \centering
    \includegraphics[width=\textwidth]{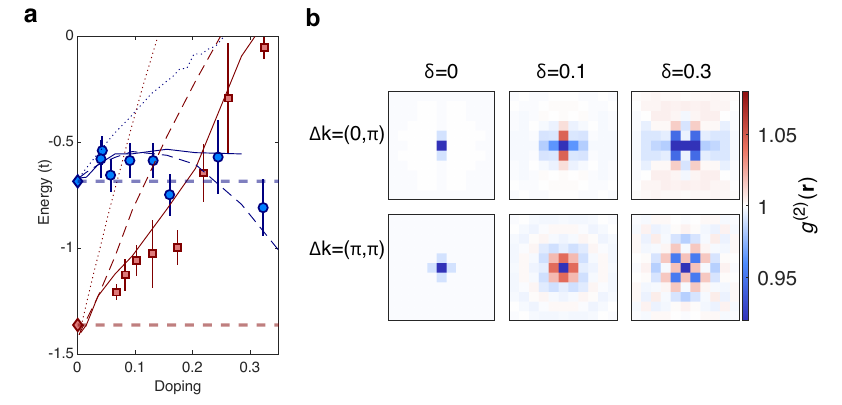}
    \caption{a. Energy shift of a magnon-Fermi polaron vs. doping. The energy of a magnon-Fermi polaron with quasimomentum $(\pi,\pi)/a$ shifts rapidly with doping (red line), while the doping dependence of a polaron with quasimomentum $(0,\pi)/a$ is much weaker. b. This behavior can be understood by considering the density correlations around the spin-flip in the polaron frame. For the first case (bottom row), the nearest-neighbor correlations are isotropic for weak doping, leading to a strong mean-field shift. For the second case (top row), the correlations are positive along one direction and negative along the other, leading to a vanishing mean-field shift. Adapted from~\cite{prichard2025observation}.}
    \label{fig:magnon_fermi_polaron}
\end{figure}

\section{Hydrodynamic transport in Fermi-Hubbard systems}
Quantum gas microscopes allow imaging microscopic transport phenomena as we have seen in Sec.~\ref{sec:magnetic_polarons}. However, they have also been useful in quantitative studies of hydrodynamic spin and charge transport. Hydrodynamics is an effective field theory that describes the collective dynamics of a many-body system in terms of the flow of conserved quantities (e.g., particle or energy density) during the approach to thermal equilibrium. It emerges at long length scales and time scales.

\label{sec:transport}
\subsection{Charge transport}

\begin{figure}[t]
    \centering
    \includegraphics[width=\textwidth]{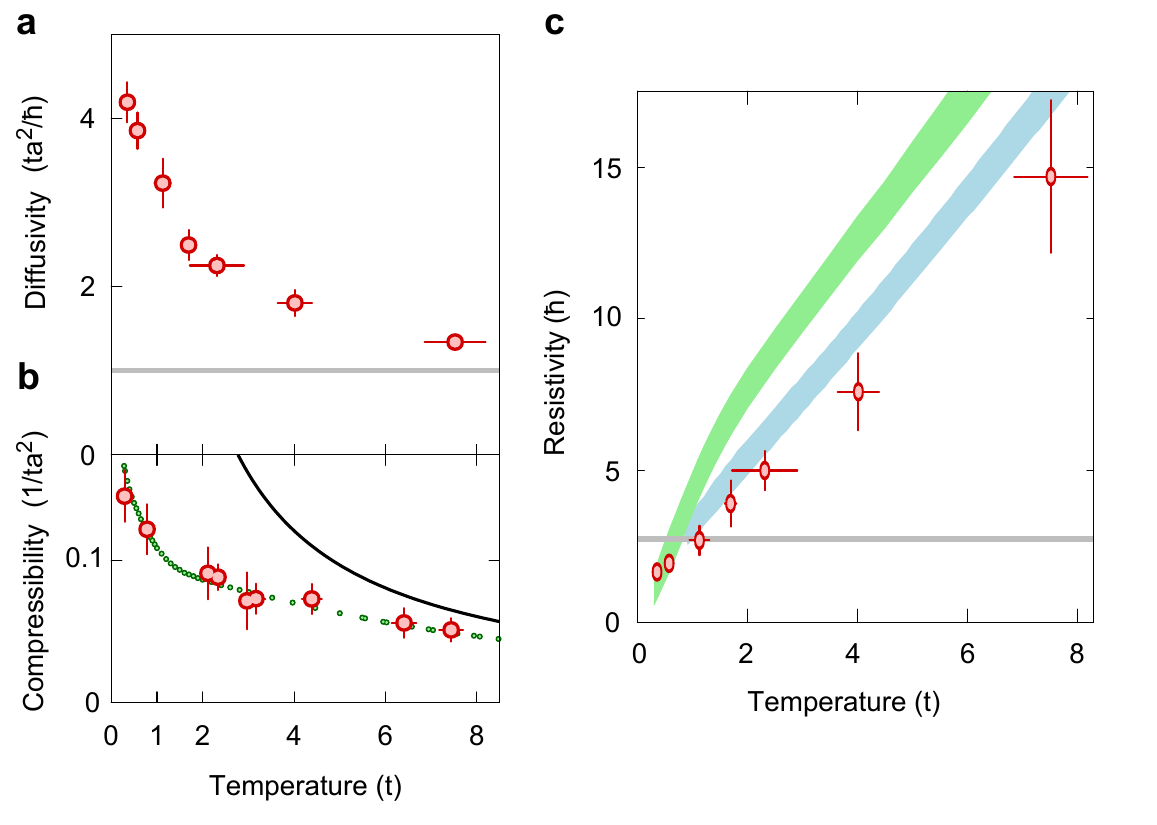}
    \caption{a. Charge diffusivity vs. temperature in the 2D Fermi-Hubbard model in the strange metal regime. b. Charge compressibility vs. temperature. c. Resistivity vs. temperature extracted from the previous two measurements using the Nernst-Einstein relation. Despite the non-trivial temperature dependence of the diffusivity and compressibility in the strange metal regime, the resistivity exhibits a $T$-linear dependence, in agreement with finite-temperature Lanczos numerics (blue). Adapted from~\cite{brown2019bad}.}
    \label{fig:charge_diffusion}
\end{figure}

The strange metal regime is an intriguing one in the cuprate phase diagram~\cite{phillips2022stranger}. It occurs in the normal phase, above the superconducting dome, and extends to quite high temperatures, up to 1000~K. It is characterized by anomalous properties that deviate from Landau's Fermi liquid theory, most importantly, a resistivity that depends linearly on temperature. 

In a Fermi liquid, conserved quantities like charge, spin or energy are carried by quasiparticles. The quasiparticle picture provides bounds on transport properties. For example, according to the Drude model, the electrical resistivity in a ``good" metal (described by quasiparticles) is linearly related to the collisional relaxation rate, which in turn is inversely proportional to the mean free path. As the metal is heated, its mean free path gets shorter and the resistivity increases. However, the resistivity is expected to eventually saturate, since it does not make sense within a quasiparticle picture for the mean free path to get shorter than the interparticle or lattice spacings. This is called the Mott-Ioffe-Regel (MIR) limit on the resistivity. Another prediction of Fermi liquid theory, in combination with Boltzmann theory, relates to the behavior of the resistivity at low temperatures. In a real material, electrical resistivity can arise because of many effects, including disorder, phonons or interactions. Each has a different dependence on temperature $T$, with interactions leading to a resistivity that scales at $T^2$. 

Many correlated materials do not follow these predictions, indicating a breakdown of the quasiparticle picture. For example, the resistivity does not exhibit saturation at the MIR limit in materials known as ``bad" metals. Moreover, the resistivity can exhibit a poorly understood linear-in-$T$ behavior, as in the strange metal regime of the cuprates. In the cuprates, it is debated whether the strange metallicity is due to strong correlation or phonon effects. This motivates asking the question of whether the Fermi-Hubbard model can host a strange metal regime. Cold atom Fermi-Hubbard systems are almost disorder-free and do not have phonons since an optical lattice is rigid. This allows for a clean study of the resistivity arising from strong correlation effects.

The Princeton group measured the temperature-dependent resistivity in a 2D Fermi-Hubbard system in the strongly interacting regime ($U/t\sim 7$) and for doping of $\sim 18\%$, parameters corresponding to the strange metal regime in the hole-doped cuprates~\cite{brown2019bad}. A direct measurement of the bulk resistivity $\rho$ in these systems is challenging. Instead, the experimentalists relied on the Nernst-Einstein relation $\sigma_c = 1/\rho_c = \chi_c D_c$, which relates the conductivity $\sigma_c$ to the charge compressibility $\chi_c = (\partial n/\partial \mu)_T$ and the charge diffusivity $D_c$. The measured temperature dependence of these three quantities is shown in Fig.~\ref{fig:charge_diffusion}. They extracted the compressibility at that particular doping from the derivative of the density profile in a harmonically trapped gas. To extract the diffusivity, they used a DMD to introduce a charge density wave into a uniformly trapped, doped Hubbard system. By tracking the decay of the charge density wave amplitude for waves of different wavelengths, they were able to extract $D_c$ using a hydrodynamic model. Interestingly, both $\chi_c$ and $D_c$ exhibit a non-trivial dependence on temperature, but $\rho_c$ is found to be linear over a large temperature range from $T/t\sim 0.3$ to $T/t\sim 8$. The experimental results have been confirmed with finite-temperature Lanczos and determinantal quantum Monte Carlo calculations.

It is important to note here that the lowest temperatures explored in the experiment (in units of the tunneling) are near the upper range of the temperatures where strange metallicity is studied in the cuprates, so the observation of linear-in-$T$ resistivity is really just a statement about a phenomenon that occurs in the Fermi-Hubbard model. In most real materials like the cuprates, the temperature-dependence of the resistivity is due to the diffusivity, while the temperature dependence of the compressibility has saturated at low temperatures. The opposite regime of very high temperatures is well-understood, as the diffusivity saturates, while the compressibility is known to scale as $1/T$ at high temperatures. The interesting observation in the experiment is that there is an intermediate temperature range where the compressibility and diffusivity conspire together to give rise the linear-in-$T$ behavior. The resistivity is also observed to violate the MIR limit, but that is so not so surprising given the expected $1/T$ dependence of the compressibility at very high temperatures.

\subsection{Spin transport}

\begin{figure}[t]
    \centering
    \includegraphics[width=0.8\textwidth]{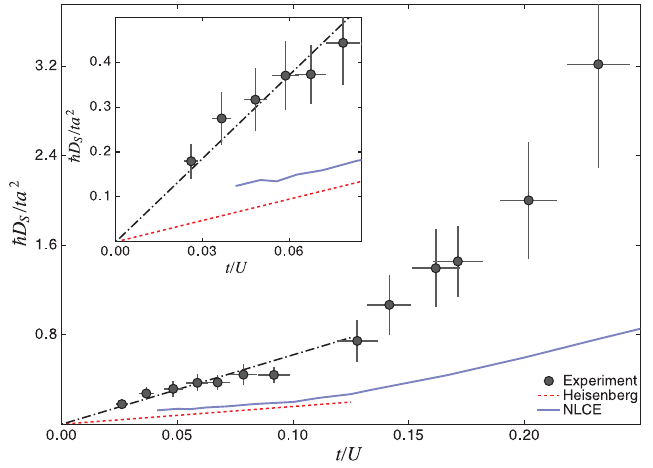}
    \caption{Spin diffusivity vs. $t/U$ in a 2D, half-filled Fermi-Hubbard system. Adapted from~\cite{nichols2019spin}.}
    \label{fig:spin_diffusivity}
\end{figure}

Stepping back from doped systems back to half-filled systems, charge transport is suppressed in the Mott insulator. Nevertheless, spin transport is still possible through superexchange processes. A similar Einstein relation to the one discussed in the previous section applies. Here $\sigma_s = \chi_s D_s$, where $\sigma_s$ is the spin conductivity, $\chi_s$ is the spin susceptibility and $D_s$ is the spin diffusivity. The MIT group studied spin transport in the Mott insulator by using a magnetic field gradient to prepare an initial state where the two spin components are initially spatially separated~\cite{nichols2019spin}. After removing the gradient, the spins diffuse into each other. The extracted spin diffusivity is shown in Fig.~\ref{fig:spin_diffusivity}. 

In a quasiparticle picture, the quantum limit for diffusion (MIR limit) is obtained by setting the mean-free path $l$ to be on order of the interparticle spacing $1/k_F$, where $k_F$ is the Fermi wavenumber. The diffusivity $D$ is $\sim vl$, where the average speed of the quasiparticles is $v\sim\hbar k_F/m$. This gives the quantum unit of diffusion, $D_0\sim \hbar/m$, where $m$ is the effective mass of the quasiparticles. For a quasiparticle system in the tight binding limit, $m \sim \hbar^2 /ta^2$. Therefore, the natural scale for diffusion is $D_0 \sim t a^2/\hbar$. In a weakly interacting system, the quasiparticles carry both spin and charge, so this scale applies to both. However, the quasiparticle picture breaks down in the Mott insulator, where spin can be transported through superexchange even when charge transport is absent. This explains the observed linear dependence of the spin diffusivity (in units of $D_0$) on $t/U$, since the effective mass is not set by $t$ but by $J = 4t^2/U$. 

An interesting finding from the measurements is that the spin diffusivity is larger than predictions from the Heisenberg model, presumably because of quantum and thermal doublon-hole fluctuations in the full Hubbard model. Furthermore, state-of-the-art numerical linked cluster expansion (NLCE) calculations are unable to reproduce the experimental data. This illustrates the challenge faced by numerical simulations of dynamics in the Hubbard model. Linear response quantities like transport coefficients can be related through Kubo relations to unequal-time correlation functions. Imaginary-time correlation functions can be generated using quantum Monte Carlo techniques for moderately large system sizes, but analytic continuation to real time is very challenging. On the other hand, techniques like NLCE do not have to deal with this issue, but are limited to small system sizes and short times. Correlations corresponding to times of order $1/t$ can be captured in the numerics, but not those for longer times of order $1/J$, relevant for spin transport.

\section{Attractive Hubbard systems}

So far, we have focused exclusively on the repulsive Hubbard model $(U>0)$, motivated by connections to high-$T_c$ superconductivity. However, the attractive Hubbard model ($U<0$) is also interesting, since it allows studying superconductivity on a lattice in a simpler setting~\cite{micnas1990superconductivity}. For example, unlike the repulsive case, quantum Monte Carlo calculations of the spin-balanced attractive Hubbard model do not suffer from the fermion sign problem at any filling. Furthermore, the Cooper pairing mechanism is much more transparent for attractive interactions. Despite the fact that the attractive model is better understood, it is still very rich, as it describes superconductivity in a strongly interacting regime where Bardeen-Cooper-Schrieffer (BCS) theory is generically inapplicable.

\begin{figure}[t]
    \centering
    \includegraphics[width=0.6\textwidth]{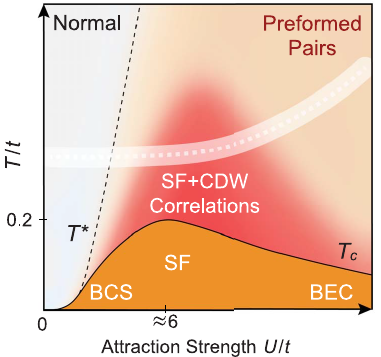}
    \caption{Phase diagram of the attractive Hubbard model at $n\approx0.8$. Below $T_c$, the system is a superfluid, with a crossover from BCS to BEC superfluidity vs. interaction strength $U/t$. The peak critical temperature occurs for intermediate interaction strength. Below $T^*$ and above $T_c$, the system is in a pseudogap regime, with a gap in the spectral function near the Fermi surface, despite the system being in a normal phase. The two temperature scales converge in the BCS regime. Adapted from~\cite{hartke2023direct}.}
    \label{fig:attractive_Hubbard_phasediagram}
\end{figure}

A cartoon phase diagram of the attractive Hubbard model in the temperature-interaction plane at a generic filling is shown in Fig.~\ref{fig:attractive_Hubbard_phasediagram} . Below a critical temperature $T_c$, there is an $s$-wave superconductor for any $U/t$. Similar to continuum attractive gases that have been extensively studied with cold atoms, there is a smooth crossover from a BCS superconductor, where pairing occurs in momentum space between fermions of opposite spin and momentum, at small $U/t$, to a Bose-Einstein condensate (BEC) of tightly-bound real-space pairs for large $U/t$. The critical temperature is maximized in the intermediate regime where the pair size and interparticle spacing are comparable. Furthermore, away from the BCS limit, there is a gap in the spectral function near the Fermi surface below a characteristic temperature $T^*$ that is above $T_c$, which defines a pseudogap regime. These features are shared with the cuprates, although the symmetry of the Cooper pairs is different.

Interestingly, the attractive Fermi-Hubbard Hamiltonian on a bipartite lattice (like the square lattice) is related to the repulsive one through a symmetry~\cite{ho2009quantum}. Consider the repulsive Hubbard model in the grand canonical ensemble and in the presence of a Zeeman field $h$. The Hamiltonian is
\begin{equation}
\hat{H}(U,\bar\mu,h)=\hat{H}_0 - \bar\mu \sum_i (n_{i\uparrow}+n_{i\downarrow})- h\sum_i (n_{i\uparrow}-n_{i\downarrow}),
\end{equation}
where $H_0$ is the Hubbard Hamiltonian in Eq.~\ref{eq:hubbard_model}, $h=\frac{1}{2}({\mu_\uparrow-\mu_\downarrow})$, $\bar\mu = \frac{1}{2}({\mu_\uparrow+\mu_\downarrow})$ and $\mu_\uparrow (\mu_\downarrow)$ are the chemical potentials for the up (down) spins respectively. Under the partial particle-hole transformation $\Lambda$,

\begin{equation}
\Lambda c_{i\uparrow} \Lambda^{\dagger} = c_{i\uparrow},~~~
\Lambda c_{i\downarrow} \Lambda^{\dagger} = (-1)^{i_x + i_y}c^{\dagger}_{i\downarrow},
\end{equation}
the Hamiltonian transforms as
\begin{equation}
\Lambda H(U,\bar\mu,h)\Lambda^\dagger = H\left(-U,h-\frac{U}{2},\bar\mu - \frac{U}{2}\right) - (\bar\mu - h)N,
\end{equation}
with $N$ the particle number. Under this transformation, it is clear that doping and spin-imbalance swap roles. For example, the doped, repulsive Hubbard model at zero field maps to the attractive model with the same absolute value of the interaction at half-filling, but in the presence of spin-imbalance.\footnote{At half-filling, the chemical potential of the repulsive (attractive) system is $U/2$ ($-U/2$).} This means that the spin-imbalanced, half-filled attractive model has a fermion sign problem, and that experimentally studying one system tells you about the other!

\subsection{Fermion pairing and charge density waves}

It immediately follows from the previous discussion that the attractive-repulsive mapping leads to a mapping of the phases of the two models onto each other. For example, the Mott phase for positive $U$ corresponds to a phase of preformed pairs at negative $U$. The pairing manifests as on-site doublons in the large $U$ limit. On the other hand, for lower interactions, the pairs become non-local, but the existence of a many-body pairing gap can be inferred from suppression of the spin susceptibility, since pairs do not contribute to the magnetization of the system. This suppression has been observed in microscope experiments by measuring the two-point magnetization correlation function~\cite{hartke2023direct}. According to the fluctuation-dissipation theorem, the integral of this quantity over all site displacements is directly proportional to the spin susceptibility.

\begin{figure}[t]
    \centering
    \includegraphics[width=\textwidth]{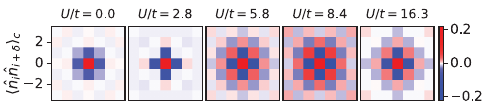}
    \caption{Measured evolution of two-point density correlations in the attractive Hubbard model vs. interaction strength $U/t$ at $n\approx0.8$. Adapted from~\cite{hartke2023direct}.}
    \label{fig:cdw_correlation}
\end{figure}

In the repulsive model, the Mott insulator develops antiferromagnetic correlations with SU(2) spin symmetry below the superexchange temperature. For negative $U$, the $x,y$ spin correlations map onto pairing correlations of the superfluid, while the $z$ spin correlations map onto charge density wave correlations. There is a degeneracy between the two types of order exactly at half-filling. At any other filling, the pairing correlations are favored and become long-range below a non-zero critical temperature.\footnote{The superconducting phase maps onto the canted antiferromagnetic phase we encountered in Sec.~\ref{sec:antiferromagnets}} The Princeton and MIT groups have measured site-resolved density correlations in attractive Hubbard systems~\cite{hartke2023direct,mitra2017quantum}. Fig.~\ref{fig:cdw_correlation} shows charge density correlations, given by $\langle \hat{n}_i {n}_j\rangle - \langle \hat{n}_i \rangle\langle{n}_j\rangle$, measured at $n\approx 0.8$. Multi-point correlations were also used to reveal polaronic disturbances in the charge density wave strength in the vicinity of single spins, closely related to magnetic polarons in the repulsive model~\cite{hartke2023direct}.

\subsection{Photoemission spectroscopy}

Spectral probes of Hubbard systems can provide complementary information to real-space correlation measurements. In the cuprates, angle-resolved photoemission spectroscopy (ARPES) has been used to confirm the theoretically predicted $d$-wave symmetry of the superconducting gap~\cite{damascelli2003}. It has also been used to study the pseudogap regime, revealing for example the existence of disconnected segments of the Fermi surface called Fermi arcs. ARPES relies on the photoelectric effect. An incoming photon with controlled energy and momentum ejects an electron from the material. The kinetic energy and emission angle of the electron are detected. Energy and momentum conservation are then used to extract the dispersion of single-particle excitations in the system.

\begin{figure}[t]
    \centering
    \includegraphics[width=\textwidth]{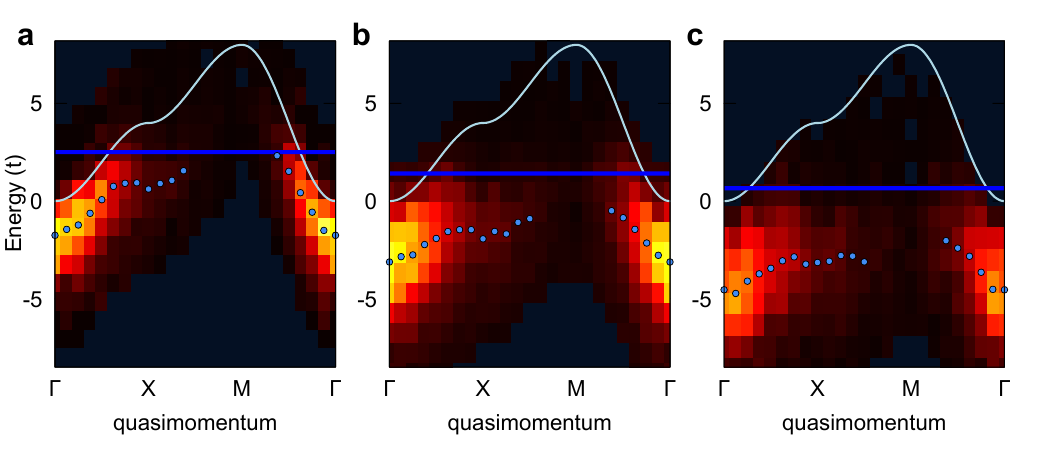}
    \caption{Measured evolution of ARPES spectra in an attractive Fermi-Hubbard system with interaction strength $U/t$. a. $U/t\approx 4$, b. $U/t\approx 6$, c. $U/t\approx 7.5$. The chemical potential is indicated by the horizontal blue line. The gap grows across the BCS to BEC crossover, as pairing transitions from many-body to two-body. Adapted from~\cite{brown2020angle}.}
    \label{fig:arpes}
\end{figure}

In cold atoms, photoemission spectroscopy was first introduced to study the excitation spectra of continuum Fermi gases with strong attractive interactions near a Feshbach resonance~\cite{stewart2008using,feld2011observation}, and later ported to Fermi-Hubbard systems by the Princeton group~\cite{brown2020angle}. In these systems, two spin components interact strongly with each other, and a radiofrequency photon with a well-defined energy is used to transfer one of the spin components to a third non-interacting state, effectively ``ejecting" it from the system. Radiofrequency photons have an energy that is appropriate for creating excitations in ultracold gases. However, their wavelength is very large compared to interparticle spacings, so they essentially transfer no momentum. Therefore, measuring the momentum of the atoms in the final non-interacting state directly gives the momentum of the quasiparticles created in the many-body system, while the known dispersion of the final state allows the extraction of their energy. In the Princeton experiment, the quasi-momentum of the final state was extracted using bandmapping followed by matter-wave focusing, where the atoms are imaged after a quarter-period oscillation in a harmonic trap, which maps momentum space to real space. Fig.~\ref{fig:arpes} shows the dispersions they measured in an attractive Hubbard system in the normal phase for different interactions across the BCS-BEC crossover. As expected, the gap at the Fermi surface grows with $U/t$, as the character of the pairing evolves from many-body to two-body. 

While this experiment was a first demonstration of ARPES in Hubbard systems, the extension to studying pseudogap physics in the doped, repulsive systems is not completely straightforward. Experiments have very recently reached temperatures below the theoretical predictions of $T^*$ according to some numerical techniques like DMFT~\cite{wu2018pseudogap,chalopin2024probing,xu2025neutral}. However, one challenge is avoiding final state interactions. The Princeton experiment worked at a specific magnetic field where the scattering length of the final state with both of the two initial spin components was negligible, a crucial aspect for effective particle ejection. A more generic approach that should work for repulsive systems involves an actual ejection of atoms from the system, e.g., into a different unpopulated, energetically offset layer using photon-assisted tunneling\footnote{Photon-assisted tunneling refers to driven tunneling across lattice wells in the presence of a large dc bias. The drive bridges the energy gap and can be provided for example by modulating the lattice depth at a frequency close to the bias.}~\cite{bohrdt2018angle}.

\section{Novel lattice geometries}
Naturally, early studies of Fermi-Hubbard systems focused on cubic lattices, and later ones with microscopes, on square lattices. These lattices are the simplest to realize experimentally, and the historical connection to cuprate physics provided a strong motivation for the work. However, the Hubbard model can be studied on different lattice geometries, which model different types of correlated materials in the solid state and exhibit rich phenomena, e.g., frustration, topological band structures and flat bands, that are absent in cubic and square lattices.

\subsection{Programmable lattices} 
Changing the lattice geometry presents a particular experimental challenge for microscope experiments, which involve complex optical setups that are difficult to modify once they are constructed. As discussed in Sec.~\ref{sec:quantumg_gas_microscopy}, compared to other ultracold lattice experiments, microscope experiments usually utilize high-power pinning lattices and advanced laser cooling schemes during imaging, which are tailored for a particular lattice setup. Therefore, until recently, almost all microscopes worked with square lattices, with the exception of microscopes at RIKEN, U. Virginia and ENS, which were specifically designed for triangular lattice geometries~\cite{yang2021site,yamamoto2020single,jongh2025quantum}.

\begin{figure}[t]
    \centering
    \includegraphics[width=\textwidth]{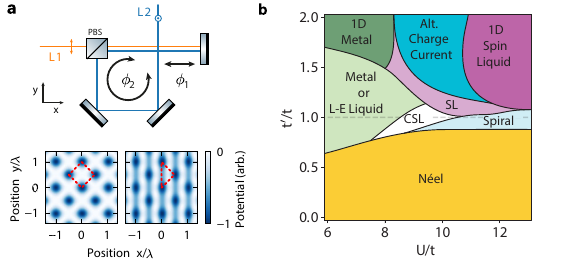}
    \caption{a. Programmable lattice introduced by the MPQ group. b. Possible phase diagram of the triangular Fermi-Hubbard model at half-filling. Adapted from~\cite{wei2023observation,szasz2021phase}.}
    \label{fig:programmable_lattice}
\end{figure}

An important advance in quantum gas microscopy was the introduction of programmable optical lattices in 2023, building on pioneering work from the ETH group~\cite{tarruell2012creating,yan2022two,xu2023doping,wei2023observation,prichard2024directly}. The effective geometry of such lattices can be controlled in software rather than by physical modification of optics setups. Fig.~\ref{fig:programmable_lattice}a shows an elegant example of such a lattice developed by the MPQ group~\cite{wei2023observation}. Two non-interfering lattices are superimposed. One is a square lattice with spacing $\lambda/\sqrt{2}$ and the other is a 1D lattice with spacing $\lambda/2$, where $\lambda$ is the wavelength of the light used to created the lattices. Both lattices share the same retroreflection mirror, so they are passively registered to each other. Two control knobs determine the lattice geometry: the frequency detuning of light used to generate the two arms controls the lattice registration phase, and the intensities of the beams control the relative lattice depths. For different parameters, Hubbard models with square, triangular, honeycomb or 1D chain geometries can be realized. Additionally, lattice sites may be ``punched out" from these geometries using repulsive potentials introduced with a DMD, which allows the realization of Lieb and Kagome geometries. In most of these cases, the lattice sites are not physically arranged according to the particular geometry that is desired, but rather, the tunneling connectivity between the sites is chosen to realize the corresponding tight-binding model. Importantly, the lattices can be adiabatically transformed back to the same square lattice for pinning during fluorescence imaging.

\subsection{Triangular Hubbard systems}
Using the programmable lattice described in the previous section, a triangular lattice Fermi-Hubbard model can be realized. Starting from a square lattice with tunneling $t$, a tunneling matrix element $t'$ along one of the diagonals is introduced. The ratio $t'/t$ can be controlled at will. Fig.~\ref{fig:programmable_lattice}b shows a cartoon phase diagram for the triangular Hubbard model with anisotropic tunneling at half-filling, which exhibits many interesting phases~\cite{szasz2021phase}. For $t'=0$, we have a square lattice, and the ground state is a N\'eel antiferromagnet for any $U/t$. This remains the case even when an appreciable $t'$ is introduced. For the isotropic triangular lattice with $t'/t=1$, the system is metallic for small $U/t$, and a spiral antiferromagnetic Mott insulator at large $U/t$. At intermediate $U/t$, the system is a non-magnetic insulator whose nature is still debated, with a chiral quantum spin liquid as one likely possibility~\cite{Szasz2020}. It is clear that the model is quite rich even at half-filling. With doping, the emergence of various types of unconventional superconductivity has been theoretically predicted~\cite{venderley2019density,zhu2022doped,zampronio2023chiral}. 

\subsubsection{Spiral antiferromagnets}

\begin{figure}[t]
    \centering
    \includegraphics[width=\textwidth]{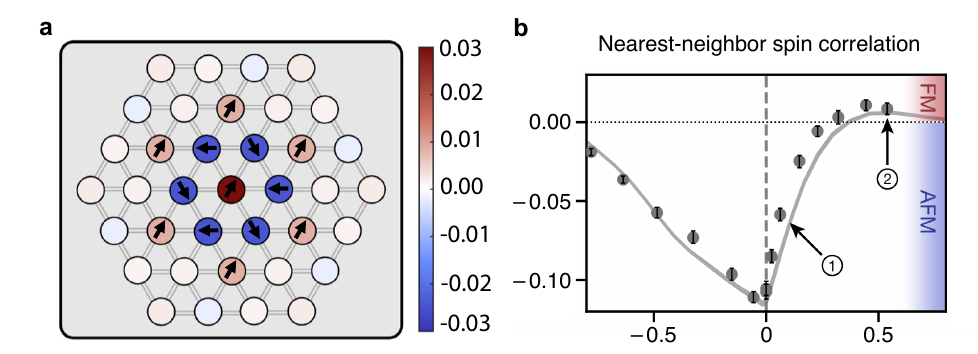}
    \caption{a. Measured two-point spin correlation function in a half-filled triangular Fermi-Hubbard system, with a classical spiral antiferromagnet pattern overlaid. b. Measured nearest-neighbor spin-correlation function in a triangular Hubbard system vs. doping. Interestingly, the correlations turn ferromagnetic for sufficiently large particle doping. Adapted from~\cite{xu2023doping}.}
    \label{fig:spiral_AFM}
\end{figure}

Motivated by the richness of this triangular Hubbard model, multiple groups have studied it with quantum gas microscopes. The U. Virginia and Harvard groups focused on physics at half-filling, observing Mott insulators with spin correlations consistent with spiral antiferromagnetic order~\cite{mongkolkiattichai2022quantum,xu2023doping}. Like the square lattice, the effective low-energy Hamiltonian for the triangular lattice at half-filling and large $U/t$ is a Heisenberg model. Antiferromagnetically coupled spins on a triangular plaquette are frustrated. This led Philip Anderson to speculate in 1973 that the quantum Heisenberg model has a spin liquid ground state~\cite{anderson1973resonating}. Later work showed this to be incorrect. Instead, the Heisenberg model does exhibit a ground state with long-range magnetic order, where the spins arrange in the plane with neighboring spins oriented at 120 degrees relative to each other. Fig.~\ref{fig:spiral_AFM}a shows the two-point correlations measured in this antiferromagnet using quantum gas microscopy, with a classical spiral pattern overlaid on it. As suggested by the classical pattern, nearest-neighbor spins are indeed found to be anti-correlated, while next-nearest-neighbor ones are positively correlated.

\subsubsection{Kinetic magnetism}

The Harvard group also studied the evolution of two-point spin correlations as the system is doped away from half-filling, shown in Fig.~\ref{fig:spiral_AFM}b~\cite{xu2023doping}. Two striking features immediately stood out in their data. First, there is an asymmetry in the correlations above and below half-filling. This is not particularly surprising, because the triangular Hubbard model lacks the particle-hole symmetry of the square lattice model. Compared to the square lattice, the correlations are enhanced below half-filling, but suppressed above half-filling. More surprisingly, the correlations turned ferromagnetic above $20\%$ particle doping. 

The latter observation was particularly intriguing, as it had possible connections to physics being studied around the same time with moir{\'e} materials~\cite{tang2020simulation,ciorciaro2023kinetic}. These are obtained by stacking 2D materials on top of each other with a controlled angle between the lattices. This gives rise to a moir{\'e} lattice whose geometry is often triangular or honeycomb. In particular, experiments with bilayer transition metal dichalcogenides, which are believed to be reasonably well-described by a triangular Hubbard model with large $U/t$, revealed interesting behavior in measurements of the spin susceptibility. The susceptibility indicated a transition between antiferromagnetic and ferromagnetic behavior as a gate was used to dope the system either with holes or electrons respectively. 

The results of both the moir{\'e} material and the cold atom experiments have been understood in terms of kinetic magnetism, a form of magnetism very different from the superexchange mediated magnetism of the square lattice~\cite{Haerter2005,zhang2018,morera2023high,davydova2023itinerant,samajdar2024nagaoka,schlomer2024kinetic}. Although the triangular incarnation of kinetic magnetism is a relatively new one, this type of magnetism in the Hubbard model actually dates back to 1966. In that year, Yosuke Nagaoka proved a rigorous theorem about the ground state of the Fermi-Hubbard model~\cite{nagaoka1966ferromagnetism}. For a class of bipartite lattices, he showed that in the limit of infinite $U/t$ and a single dopant (hole or particle), the ground state of the model is fully ferromagnetic. This surprising result, known as Nagaoka ferromagnetism, is a consequence of the constructive interference of different trajectories the hole can take through the lattice when it moves in a spin-polarized environment.\footnote{In an antiferromagnetic background, the different trajectories are distinguishable and do not interfere.} This enhanced delocalization lowers the kinetic energy of the hole, favoring ferromagnetism. Superexchange vanishes in the limit under consideration, so the magnetism is purely of a kinetic nature. Unfortunately, the theorem only holds in these idealized limits of vanishing doping and infinite interactions, which has hindered an observation in any real material. 

\begin{figure}[t]
    \centering
    \includegraphics[width=\textwidth]{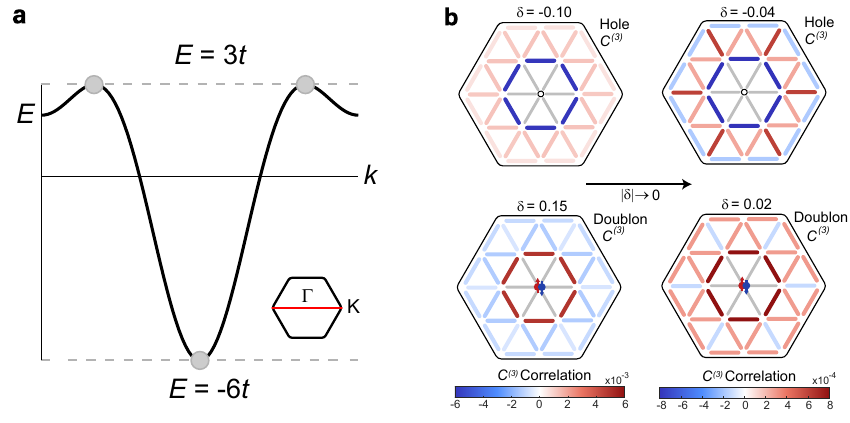}
    \caption{a. Cut through the band structure of a triangular lattice. b. Measured spin correlations around hole (top row) and particle (bottom row) dopants in a triangular Hubbard system, indicating the formation of Nagaoka-like polarons. Adapted from~\cite{prichard2024directly}.}
    \label{fig:triangular_band}
\end{figure}

Interestingly, Nagaoka magnetism turns out to be much more robust on the triangular lattice, and can persist to finite $U/t$ and doping. This is related to the phenomenon of kinetic frustration in this lattice. The band structure for a particle in an otherwise empty triangular lattice has its minimum at zero quasi-momentum, where the energy is lowered by $6t$, assuming the usual positive sign of $t$ (Fig.~\ref{fig:triangular_band}a). In contrast, a hole in a spin polarized lattice hops with the opposite sign.\footnote{This can be seen by rewriting the kinetic energy operator of the Hubbard Hamiltonian in terms of hole operators $h^\dagger_i = c_i$. Anti-commuting the hole operators to obtain a normally ordered kinetic energy term flips the sign of the tunneling.} This flips the dispersion, and the minima are now at non-zero momentum. The energy of the hole is only lowered by 3$t$, so the hole is said to be kinetically frustrated compared to the particle. This can be traced back to the fact that there is an odd number of bonds in a triangular plaquette, which leads to destructive interference of hole trajectories in a spin-polarized background. The interference is removed in an antiferromagnetic background, which introduces which-path information. Hence, the antiferromagnetic background releases the hole frustration energy, making antiferromagnetism favorable.\footnote{This is called Haerter-Shastry antiferromagnetism, named after two physicists who studied the motion of single hole in an infinite-$U$ Hubbard model on a frustrated triangular lattice~\cite{Haerter2005}.} On the other hand, a doublon dopant behaves as a particle, and favors ferromagnetism.

In the limit of very light doping, the dopants act as impurities which are polaronically dressed by magnetic excitations. These Nagaoka-like polarons have been observed by the Princeton and Harvard groups using connected three-point correlations functions~\cite{prichard2024directly,lebrat2024observation}. Fig.~\ref{fig:triangular_band}b shows the spin correlations measured around hole or particle dopants in the triangular lattice, which show local antiferromagnetic or ferromagnetic ``bubbles" around the dopants, respectively. These polarons, which are essentially a bound state of a dopant with a magnon, are the fundamental building of blocks of kinetic magnetism in the triangular Hubbard model. They have also been observed in moir{\'e} bilayers as magnetization plateaus in the response to an applied magnetic field~\cite{tao2024observation}.

Since kinetic magnetism occurs at the $t$ scale rather than the $J$ scale, it survives to rather high temperatures. More complex polarons have also been predicted in this system, where multiple holes and magnons bind together~\cite{morera2024attraction}. The overall energy scale for these polarons remains $t$, although the prefactor is smaller than the polaron studied in the cold atom experiments. As a result, these polarons may provide a mechanism for hole pairing and high-temperature superconductivity~\cite{schrieffer1988spin}. However, the temperatures achieved in the experiments to date were too high to detect hole pairing.

\subsection{Mixed dimensional systems}

\begin{figure}[t]
    \centering
    \includegraphics[width=\textwidth]{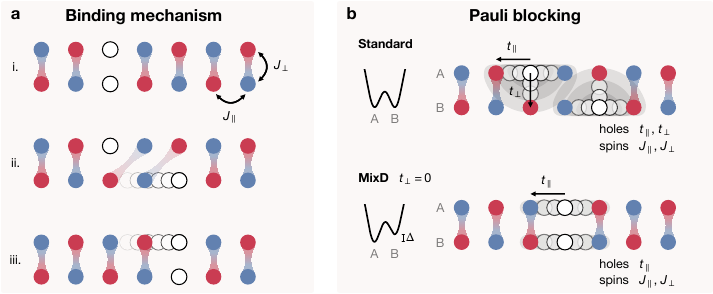}
    \caption{a. Hole binding mechanism in a lightly doped Fermi-Hubbard ladder. b. In a normal ladder, Pauli repulsion between holes competes with the binding mechanism. This competition can be eliminated by moving to a mixed dimensional system where hole tunneling only occurs along the ladder's legs, while antiferromagnetic superexchange is strong along the rungs. Adapted from~\cite{hirthe2023magnetically}.}
    \label{fig:mixed_dimensions}
\end{figure}

Technical challenges in cooling cold atom Fermi-Hubbard systems have inspired exploration of other approaches to enhance the temperatures at which hole pairing occurs. One successful approach demonstrated by the MPQ group is based on ``mixed dimensions"\footnote{At the time of these experiments, the lowest temperatures achieved on the square lattice were $\sim 0.5J$, with hole pairing expected at temperatures about an order of magnitude lower.}~\cite{bohrdt2022strong,hirthe2023magnetically}. These experiments studied Mott insulators with light hole doping on a ladder system. If the superexchange is arranged to be much stronger along the rung direction than the leg direction, spin singlets form along the rungs. As can be seen in Fig.~\ref{fig:mixed_dimensions}a, this gives rise to a hole binding mechanism along rungs, since breaking a hole pair on a rung displaces the spin singlets with an energy cost of order $J$. However, the hole pairing mechanism competes with Pauli blocking, which favors the holes staying away from each other so they can delocalize along rungs. The solution is to introduce a large energy offset between the legs, which inhibits hole tunneling along the rungs, resulting in 1D hole dynamics\footnote{Note that the energy offset switches off the tunneling along rungs, but not the superexchange.} (Fig.~\ref{fig:mixed_dimensions}b). This suppresses the competing delocalization energy, and leads to a hole binding energy that is estimated to be $\sim 0.8J$. The group indeed observed hole bunching along rungs~\cite{hirthe2023magnetically}.

The group also extended this work to a 2D system with the energy offsets introduced on alternating rows. This allowed them to observe a precursor to a stripe phase, where hole dopants exhibit bunching correlations and an increased probability for forming extended charge stripe structures~\cite{bourgund2025formation}. It is interesting to note connections to the bilayer nickelates, where high-temperature superconductivity has been recently observed~\cite{sun2023signatures}. The dynamics of these materials is also believed to be mixed-dimensional in nature.

\subsection{Lieb lattices}

\begin{figure}[t]
    \centering
    \includegraphics[width=\textwidth]{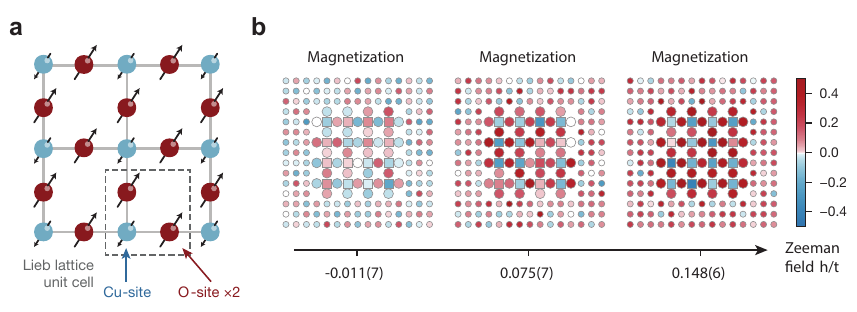}
    \caption{a. Lieb lattice geometry. b. A half-filled Lieb lattice is realized in the experiment, surrounded by a square lattice. In the presence of an effective field created by introducing spin-imbalance, the Lieb region exhibits a ferrimagnetic susceptibility. The Cu-sites and O-sites have anti-aligned spins, and the region has an overall magnetization since there are twice as many O-sites in a unit cell as Cu-sites. Adapted from~\cite{lebrat2024ferrimagnetism}.}
    \label{fig:lieb}
\end{figure}

Another interesting lattice geometry that has been explored with quantum gas microcopy is the Lieb lattice. The Lieb lattice is a decorated square lattice with three sites per unit cell, as shown in Fig.~\ref{fig:lieb}. It is also known as the CuO lattice because of its similarity to the lattice structure of the cuprate planes in high-$T_c$ superconductors, with copper sites at the corners of the square plaquette ($A$ sites), and oxygen sites located at the centers of the edges ($B$ and $C$ sites). The single-particle band structure of the tight-binding model exhibits a flat band sandwiched between two dispersive bands with Dirac cones. The flat band is the result of destructive interference of the hopping amplitudes from the $B$ and $C$ sites onto the $A$ sites, leading to the emergence of localized states on the $B-C$ sites.

The Lieb lattice is bipartite, with its $B-C$ sublattice having twice as many sites as the $A$ sublattice. According to a rigorous theorem proved by Lieb, this means that the ground state of the Hubbard model on the Lieb lattice has non-zero total spin at half-filling for any interaction strength~\cite{lieb1989two}. This is easy to understand in the large $U$ limit, where the system forms a Mott insulator, and a description in terms of an antiferromagnetic Heisenberg model is appropriate. Since the magnetic moments on the two sublattices do not balance, the antiferromagnet has a net magnetization, a phenomenon known as ferrimagnetism. In the opposite extreme of vanishing $U$, the flat band corresponds to a very high degeneracy of states. Any small repulsive interaction lifts this degeneracy, and only states with maximum total spin remain as ground states.

The Harvard group demonstrated various aspects of this physics~\cite{lebrat2024ferrimagnetism}. By differentiating the measured local density profile with respect to the chemical potential, they showed a peak in the local compressibility on the $B-C$ sites near half-filling, as expected from crossing a flat band. Spin conservation in cold atom systems does not permit observing ferrimagnetism through spontaneous symmetry breaking, but they were still able to measure a ferrimagnetic susceptibility by applying a small effective Zeeman field (Fig.~\ref{fig:lieb}b).

\section{Novel interactions}

All the realizations of the Fermi-Hubbard model we have discussed to this point exhibit purely on-site interactions to excellent approximation. This is because van der Waals interactions between ground state atoms on neighboring sites of a lattice are negligible compared to the tunneling matrix element. There is a strong motivation to investigate ``extended" Hubbard models with off-site interactions. Quantum simulations with off-site interactions would capture the behavior of certain electronic materials like moir{\'e} materials more accurately. Furthermore, off-site interactions can lead to the emergence of interesting phases of matter with spatial structure, such as fractional Mott insulators which possess long-range density order despite the unit cell having a fractional filling. Finally, off-site interactions can lead to frustration even on non-frustrated geometries. For example, the square lattice Heisenberg model becomes frustrated if there is a competition between the spin alignments favored by the nearest-neighbor and next-nearest-neighbor exchange couplings. 

The microscopy of extended Hubbard models is being pursued actively in various platforms. One approach involves using Rydberg states of atoms. Atoms in Rydberg states are highly polarizable, and interact strongly even at distances of a few microns. In quantum gas microscopes, this has enabled the observation of mesoscopic crystalline structures of Rydberg excitations~\cite{schuass2012observation}. Most experiments have been restricted to the ``frozen" gas regime, where the atoms do not tunnel. This is because the lifetime of an atom in a Rydberg state is typically on the order of a hundred microseconds, short compared to tunneling times. Furthermore, the strong Rydberg interactions completely overwhelm the tunneling energy scale, only allowing for crystal states. 

Rydberg dressing provides a way around these limitations~\cite{Henkel-Pohl2010,Johnson-Rolston2010,Pupillo-Zoller2010}. By coupling to the Rydberg state off-resonantly, the atomic wavefunction is predominantly the ground-state one, but with a small Rydberg admixture which introduces a long-range interaction. Since the admixture is small, the lifetime of the dressed state is greatly enhanced and can exceed the tunneling time. The Princeton group used Rydberg dressing of a spin-polarized Fermi gas to realize a $t-V$ model~\cite{guardadosanchez2021quench}. In this model, itinerant fermions interact through a soft-core interaction, which is primarily nearest-neighbor. They observed a slowdown of the relaxation dynamics of imprinted charge-density wave states in this system due to the strong interactions. However, the system was restricted to exploring short times due to decay from the Rydberg admixture. Longer lifetimes were recently achieved by the MPQ group in an extended Bose-Hubbard model, allowing the observation of density ordering near equilibrium~\cite{weckesser2024realization}.

More promising approaches are based on quantum gases of highly magnetic atoms or polar molecules. These quantum gases can be long-lived, with lifetimes on the order of several seconds. Microscopy of these quantum gases has only been demonstrated with bosonic species of these particles, but the extension of the techniques to fermionic species for studying extended Fermi-Hubbard models should be straightforward. 

Unlike the alkali atoms we have discussed, atoms like dysprosium or erbium have multiple unpaired electrons, leading to large magnetic moments which have been utilized to study various aspects of the physics of dipolar quantum gases~\cite{chomaz2022dipolar}. The Harvard group has realized a quantum gas microscope for bosonic erbium, which has a magnetic moment of $7 \mu_B$, where $\mu_B$ is a Bohr magneton. That microscope is special in that it does not utilize a pinning lattice or an advanced laser cooling scheme for fluorescence imaging. Instead, an accordion lattice is used to increase the lattice spacing just before imaging~\cite{su2025fast}. The large spacing (2-5$~\mu$m) allows high-fidelity fast free-space imaging ($\sim2~\mu$s) in which only $\sim 15$ photons are collected. The microscope was used to study a dipolar Bose-Hubbard model, where they observed quantum phase transitions from superfluids to dipolar quantum solids~\cite{su2023dipolar}. These are phases where there is density order in the system accompanied by strong quantum fluctuations due to tunneling. 

The interactions between magnetic atoms on neighboring sites of an optical lattice are quite weak, typically on the order of tens of Hz. Much stronger interactions can be achieved with electric dipoles in polar molecules~\cite{langen2024quantum}. Degenerate gases of polar molecules have been prepared using \textit{in-situ} assembly approaches. These involve magneto-association of Feshbach molecules starting from ultracold gases of two atomic species, followed by transfer to the rovibrational ground state of the molecule. For a polar molecule, strong electric dipolar interactions can be induced by mixing rotational states of opposite parity, either using dc electric fields or microwaves. 

The Princeton group has demonstrated quantum gas microscopy of bosonic NaRb polar molecules~\cite{christakis2023probing}. By dissociating the molecule to its atomic components and then detecting the Rb atoms, they extracted the site-resolved positions of the molecules. Furthermore, they could encode a pseudospin-1/2 in the rotational states, realizing various dipolar spin models and characterizing correlation dynamics in them. Very recently, the Durham group has also extended microscopy techniques to RbCs, another bosonic molecule~\cite{mortlock2025multi}. Although these experiments were performed with pinned molecules, meaning that their quantum statistics did not matter, molecular gases can also be studied in an itinerant regime~\cite{gorshkov2011tunable}. Inelastic collisions between itinerant molecules are a concern, but the development of collisional shielding techniques has allowed strong suppression of these loss mechanisms~\cite{langen2024quantum}. 

Indeed, recent experiments with KRb have started exploring itinerant fermionic models with dipolar interactions (without microscopy)~\cite{carroll2025observation}. The realized model is a generalization of the $t-J$ model, whose Hamiltonian is given by

\begin{multline}
\hat{H} = -t\sum_{\langle i,j\rangle,\sigma}(\hat{c}_{i,\sigma}^{\dagger}\hat{c}_{j,\sigma} + \mathrm{h.c.})+U\sum_i\hat{n}_{i\uparrow}\hat{n}_{i\downarrow}\\+\sum_{i\neq j}\frac{V_{ij}}{2}\left[J_\perp \hat{\mathbf{S}}^\perp_i\cdot \hat{\mathbf{S}}^\perp_j + J_z \hat{S}^z_i \hat{S}^z_j+V\hat{n}_{i}\hat{n}_j+W(\hat{n}_{i}\hat{S}^z_j+\hat{n}_{j}\hat{S}^z_i)\right].
\end{multline}
Here, both the very large on-site interaction $U$ and a quantum Zeno effect associated with rapid on-site loss prevent double occupancies, as is the case in the $t-J$ model derived from the Fermi-Hubbard model. However, the dipolar interaction leads to additional off-site terms given by the second line, with a dipolar geometric factor $V_{ij} = (1-3\cos^2{\theta_{ij}})a^3/r_{ij}^3$. Although there is no superexchange due to vanishing double occupancies, there is a direct, anisotropic spin coupling, as well as density-density and spin-density couplings, all of which are tunable. In the usual $t-J$ model derived from the Hubbard model, $J = 4t^2/U$, so $J$ is small compared to $t$ in the regime of strong coupling where the $t-J$ model is a reasonable approximation. In contrast, the polar molecule realization of the generalized $t-J$ models allows  access to the regime where $J$ is large compared to $t$, since the parameters are controlled independently. This is interesting because it can increase the critical temperature for exotic phases (e.g., unconventional superfluids). However, current experiments on polar molecules in optical lattices have not yet achieved the same level of degeneracy as atomic gases, and further progress on cooling these gases is necessary before itinerant many-body phases can be explored in equilibrium.

\section{Entering the discovery regime}

The rapid advances in microscopy of ultracold fermions in optical lattices over the past decade have been truly amazing. From observations of magnetic polarons that had only been discussed in textbooks, to the discovery of new regimes of strange metallicity, the rich physics revealed by these experiments is deepening our understanding of many-body systems and challenging state-of-the-art numerics. However, given that most of these experiments have been performed at temperatures corresponding to about 1000~K in the solid state, it is clear that there is still much more to learn by cooling atomic Fermi gases to lower temperatures. 

The temperature of a correlated state in an optical lattice is determined by three factors: the total entropy of the Fermi gas at the end of evaporative cooling, any additional entropy introduced while preparing the correlated state due to non-adiabatic processes, and the distribution of the entropy in the inhomogeneous system. In principle, evaporative cooling does not have any fundamental limits to the temperature it can reach, which is determined by a balance between the cooling and heating rates. Heating mechanisms include light scattering from optical potentials, intensity, frequency and beam pointing noise of lasers and inelastic collisions, especially near a Feshbach resonance. Background gas collisions also lead to ``hole heating" in a Fermi gas, as removing a random fermion from a Fermi sea generates high energy excitations. The typical entropy per particle achieved at the end of evaporative cooling is $\sim1~k_B$. By comparison, the entropy increase per particle due to non-adiabaticities while ramping on the lattice are usually small, especially if care is taken to minimize mass transport. As discussed in Sec.~\ref{sec:engineering_potential}, careful shaping of the trapping potential can help redistribute entropy in the system between different regions. The lowest temperatures achieved with this approach are $\sim 0.25t$. This is around the temperature where the fermion sign problem becomes severe for quantum Monte Carlo calculations for doped systems. 

Recently, a more advanced entropy redistribution technique has been realized by the Harvard group, allowing them to reach significantly lower temperatures, where unbiased numerics cannot be performed~\cite{xu2025neutral}. This is the highly sought ``discovery regime" of quantum simulators. Of course, approximate numerics are still available, but their predictions can no longer be trusted, especially in regimes where there may be multiple competing phases. The idea implemented by the group is an old one, but its realization has been technically challenging~\cite{bernier2009cooling,lubasch2011adiabatic}. The trapping conditions were arranged to obtain a band insulator surrounded by large metallic reservoir. The large gap in the band insulator efficiently expels entropy into the gapless metal, leading to a very low entropy of $\sim 0.025~k_B$ per particle in the insulator. The insulator was then isolated from the metal by introducing a barrier potential. It was then transformed into a correlated state by splitting each lattice site into two, converting it into an antiferromagnet at a temperature of $T\sim0.05t$ for $U/t\sim8$. More interestingly, the gas could be adiabatically expanded slightly in a shallow lattice, before bringing it back into the strongly-correlated regime, introducing doping. Comparisons to an approximate numerical technique suggested a temperature below $0.1t$. In principle, a different lattice could be ramped on after the expansion. This opens up a new chapter in exploring low-temperature many-body phases of the Fermi-Hubbard model on a variety of lattice geometries. 

\section{Emerging directions}

While these lectures have focused on analog simulations of Fermi-Hubbard systems in quantum gas microscope experiments, it is worth pointing out some other related developments in the field. On the theoretical front, as these experiments enter regimes which are not theoretically understood, identifying the microscopic correlations responsible for new phenomena becomes challenging. The experimental snapshots contain a wealth of information beyond few-point correlation functions. Machine learning techniques have been introduced to analyze snapshots, e.g., for classifying experimental snapshots into theory categories using neural networks trained on simulated data~\cite{bohrdt2019classifying}. Alternatively, experimental snapshots taken in different regimes can be used to train neural networks and identify hidden orders~\cite{khatami2020visualizing}. The main challenge with this approach is the amount of data that can be generated experimentally for training.

On the experimental front, microscopy of ultracold fermions continues to expand to new species and modalities. For example, quantum gas microscopy of bosonic two-electron atoms (Yb and Sr) has been demonstrated~\cite{yamamoto2016ytterbium,miranda2017site,buob2024strontium}. Extensions to fermionic isotopes, which have already been studied without microscopy~\cite{taie2022observation,pasqualetti2024equation}, would allow detailed explorations of SU($N$) magnetism by taking advantage of the decoupling of the nuclear spin from the electronic angular momentum. Another emerging direction is using microscopy to study continuum systems, where the larger energy scales makes accessing many-body phases easier~\cite{jongh2025quantum,xiang2025insitu,yao2025measuring}. This has been applied to study spin and charge correlations in continuum strongly interacting Fermi gases near a Feshbach resonance~\cite{yao2025measuring,daix2025observing}. Future work could explore polaron correlations in spin-imbalanced gases, related to those studied in lattice systems.

One of the challenges of working with analog Fermi-Hubbard simulators in the ``discovery regime" is verification of the data they generate. How do we ensure that the measurements are not affected by unknown Hamiltonian terms or couplings to the environment? Here, the advantages of digital quantum computers, with their error detection and correction capabilities, become apparent. Digital simulations of Fermi-Hubbard models have been realized in various quantum computing platforms, but the sizes of the systems and the times over which the dynamics can be simulated are quite limited in comparison to the analog simulations described in these lectures~\cite{barends2015digital,nigmatullin2024experimental,evered2025probing}. To date, digital simulations of Fermi-Hubbard systems have been restricted to physical qubits, and extensions to logical qubits with error correction are currently out of reach. Part of the challenge lies in the large overhead involved in encoding fermions into qubits, especially in the case of Fermi-Hubbard models describing non-periodic systems such as those encountered in quantum chemistry. This has motivated proposals of fermionic neutral atom quantum computers that circumvent this overhead by working with native fermions~\cite{gonzalez2023fermionic,ott2024error,schukert2024fermion}. A qubit-based quantum computing platform, e.g., Rydberg atoms with qubits encoded in internal states, is supplemented with digital tunneling gates so that the Fermi statistics is hardwired into the computer. Experimentally, initial steps in the direction of fermionic quantum computing have been realized with quantum gas microscopes, where high-fidelity collisional gates between fermionic atoms have been demonstrated~\cite{bojovic2025high}. While full-blown fermionic quantum computation remains a long-term prospect, it is reasonable to foresee hybrid digital-analog simulations in the near future, e.g. for reading out entanglement and coherences in many-body states.

\begin{acknowledgement}
We thank Martin Zwierlein and Lawrence Cheuk for helpful comments on the manuscript. We also thank the groups of Markus Greiner, Martin Zwierlein and Immanuel Bloch for the figures used in this manuscript, and for providing a stimulating extended research community over the years. We acknowledge support by the NSF (grant no. 2110475), the David and Lucile Packard Foundation (grant no. 2016-65128) and the ONR (grant no. N00014-21-1-2646). MLP acknowledges support by the NSF Graduate Research Fellowship Program.  
\end{acknowledgement}
\bibliographystyle{spphys}
\bibliography{ref}
\end{document}